\documentclass[a4paper,11pt]{article}
\pdfoutput=1

\usepackage{jheppub}

\usepackage{graphicx}
\usepackage{amsmath}
\usepackage{amssymb}
\usepackage{amsfonts}
\usepackage{dcolumn}
\usepackage{bm}
\usepackage[dvipsnames]{xcolor}
\usepackage[utf8]{inputenc}
\usepackage{subfigure}
\usepackage{gensymb}
\usepackage{cleveref}

\usepackage[T1]{fontenc}
\usepackage{pstricks}
\usepackage{color}
\usepackage{multirow}
\usepackage{slashed}
\usepackage{mathtools}

\usepackage{feynmp}
\usepackage{feynmp-auto}
\newcommand{\N}{\mathcal{N}}

\newcommand{\UNINA}{Dipartimento di Fisica "Ettore Pancini", Università degli studi di Napoli "Federico II", \\ Complesso Univ. Monte S. Angelo, I-80126 Napoli, Italy}

\newcommand{\INFN}{INFN - Sezione di Napoli, Complesso Univ. Monte S. Angelo, I-80126 Napoli, Italy}

\newcommand{\SOTON}{Department of Physics and Astronomy, University of Southampton, SO17 1BJ Southampton, United Kingdom}

\title{\boldmath Dark Matter in the Type Ib Seesaw Model}

\author[a,b,1]{M. Chianese,\note{\url{https://orcid.org/0000-0001-8261-4441}}}
\author[c,2]{B. Fu,\note{\url{https://orcid.org/0000-0003-2270-8352}}}
\author[c,3]{S. F. King\note{\url{https://orcid.org/0000-0002-4351-7507}}}

\affiliation[a]{\UNINA}
\affiliation[b]{\INFN}
\affiliation[c]{\SOTON}

\emailAdd{marco.chianese@unina.it}
\emailAdd{B.Fu@soton.ac.uk}
\emailAdd{king@soton.ac.uk}

\abstract{
We consider a minimal type Ib seesaw model where the effective neutrino mass operator involves two different Higgs doublets, and the two right-handed neutrinos form a heavy Dirac mass. We propose a minimal dark matter extension of this model, in which the Dirac heavy neutrino is coupled to a dark Dirac fermion and a dark complex scalar field, both charged under a discrete $Z_2$ symmetry, where the lighter of the two is a dark matter candidate. Focussing on the fermionic dark matter case, we explore the parameter space of the seesaw Yukawa couplings, the neutrino portal couplings and dark scalar to dark fermion mass ratio, where correct dark matter relic abundance can be produced by the freeze-in mechanism. By considering the mixing between the standard model neutrinos and the heavy neutrino, we build a connection between the dark matter production and current laboratory experiments ranging from collider to lepton flavour violating experiments. For a GeV mass heavy neutrino, the parameters related to dark matter production are constrained by the experimental results directly and can be further tested by future experiments such as SHiP. }


\arxivnumber{2102.07780}

\begin{document}

\maketitle

\flushbottom

\section{Introduction \label{sec:intro}}

The masses of neutrinos and their mixing, evidenced by the neutrino oscillation experiments~\cite{2016NuPhB.908....1O}, is one of the open questions in particle physics and indicates the existence of new physics beyond the Standard Model (SM). Theorists have developed multiple theories to explain the origin of the neutrino masses, most of which are different realisations of the dimension-five Weinberg operator~\cite{Weinberg:1979sa}. Typical tree-level realisations of the Weinberg operator include the type I~\cite{Minkowski:1977sc,Yanagida:1979as,GellMann:1980vs,Mohapatra:1979ia}, II~\cite{Magg:1980ut,Schechter:1980gr,Wetterich:1981bx,Lazarides:1980nt,Mohapatra:1980yp,Ma:1998dx} and III~\cite{Foot:1988aq,Ma:1998dn,Ma:2002pf,Hambye:2003rt} seesaw models. However, large seesaw coupling and small right-handed (RH) neutrino mass cannot be simultaneously achieved in the traditional seesaw models, which makes them hard to test. 
For this reason, much attention has been focussed also on low scale seesaw models such as the inverse seesaw model~\cite{Mohapatra:1986bd} or the linear seesaw model~\cite{Akhmedov:1995ip,Malinsky:2005bi}, where both sorts of model are based on extensions of the right-handed neutrino sector. Loop models of neutrino mass provide further low energy alternatives~\cite{Ma:2009dk}. 

\begin{figure}[t!]
\begin{center}
\begin{fmffile}{NM}
\fmfframe(23,18)(18,18){
\begin{fmfgraph*}(200,45)
\fmflabel{$L_\beta$}{o1}
\fmflabel{}{o2}
\fmflabel{}{i2}
\fmflabel{$L_\alpha$}{i1}
\fmfv{}{v1}
\fmfv{decor.shape=cross,decor.size=8,label=$M_N$,label.angle=90}{v2}
\fmfv{}{v3}
\fmfv{label=$\Phi_1$,label.angle=90}{v4}
\fmfv{}{v5}
\fmfv{label=$\Phi_2$,label.angle=90}{v6}
\fmfleft{i1,i2}
\fmfright{o1,o2}
\fmf{fermion,tension=0.6}{i1,v1}
\fmf{fermion,label=$N_{R1}$,l.side=left}{v2,v1}
\fmf{fermion,label=$N_{R2}$,l.side=right}{v2,v3}
\fmf{fermion,tension=0.6}{o1,v3}
\fmf{phantom,tension=0.6}{i2,v4}
\fmf{phantom}{v5,v4}
\fmf{phantom}{v5,v6}
\fmf{phantom,tension=0.6}{o2,v6}
\fmf{dashes,tension=0}{v1,v4}
\fmf{phantom,tension=0}{v2,v5}
\fmf{dashes,tension=0}{v3,v6}
\end{fmfgraph*}}
\end{fmffile}
\caption{\label{fig:Ib} The type Ib seesaw mechanism involves two different Higgs doublets $\Phi_1$ and $\Phi_2$. The minimal model involves two right-handed neutrinos $N_{\mathrm{R}1}$ and $N_{\mathrm{R}2}$ which form a Dirac mass $M_N$.}
\end{center}
\end{figure}
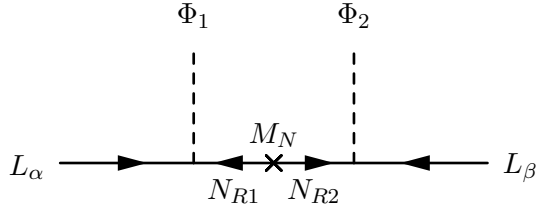

Recently a new version of the type I seesaw mechanism has been proposed called the  
type Ib seesaw mechanism~\cite{Hernandez-Garcia:2019uof} which may be just as
testable as the low scale seesaw models above while allowing just two right-handed neutrinos~\cite{King:1999mb}.
It had been pointed out a long time ago that the traditional Weinberg operator is not the only pathway to neutrino mass in models with multiple Higgs doublets~\cite{Oliver:2001eg}. The two Higgs doublet models (2HDMs) have been classified by many authors~\cite{Aoki:2009ha,Branco:2011iw,Chao:2012pt}, and the 
type Ib seesaw model~\cite{Hernandez-Garcia:2019uof} is based on the so called type II 2HDM in which one Higgs doublet couples to down type quarks and charged leptons, while the other couples to up type quarks. Usually when the type I seesaw mechanism
is combined with the 
type II 2HDM, the Weinberg operator~\cite{Weinberg:1979sa} would involve only the Higgs doublets that couple to up type quarks. The novel feature of the 
type Ib seesaw mechanism is that the effective neutrino mass operator requires both types of Higgs doublets, which couple to up and down type quarks,
while the two right-handed neutrinos form a single Dirac mass in the minimal case~\cite{Hernandez-Garcia:2019uof},
as shown in Fig.\ref{fig:Ib}.
In contrast to the traditional type I seesaw models, the type Ib seesaw model allows large seesaw couplings and relatively small value of the heavy neutrino mass simultaneously and thus is more testable than the traditional type I seesaw model, which we shall refer to as type Ia to distinguish it from type Ib. In this case the type Ib seesaw model~\cite{Hernandez-Garcia:2019uof} shares many of the general features of testability as the inverse seesaw or linear seesaw models mentioned above, however it is distinguished by the simplicity of the two right-handed neutrino sector~\cite{King:1999mb} which form a single heavy Dirac mass in the minimal case, as mentioned above,
rather than relying on extending the right-handed neutrino sector as in other low energy seesaw models. This makes the minimal type Ib seesaw model particularly well suited for studying dark matter produced via a heavy neutrino portal, as we now discuss.

Besides neutrino mass and mixing, the existence of dark matter (DM) accounting for about 25\% of the energy density of the universe~\cite{Aghanim:2018eyx} also provides an important clue of physics beyond the standard model (BSM). The relation between these intriguing phenomena has been investigated in many works~\cite{Caldwell:1993kn,Mohapatra:2002ug,Krauss:2002px,Ma:2006km,Asaka:2005an,Boehm:2006mi,Kubo:2006yx,Ma:2006fn,Hambye:2006zn,Lattanzi:2007ux,Ma:2007gq,Allahverdi:2007wt,Gu:2007ug,Sahu:2008aw,Arina:2008bb,Aoki:2008av,Ma:2008cu,Gu:2008yj,Aoki:2009vf,Gu:2010yf,Hirsch:2010ru,Esteves:2010sh,Kanemura:2011vm,Lindner:2011it,JosseMichaux:2011ba,Schmidt:2012yg,Borah:2012qr,Farzan:2012sa,Chao:2012mx,Gustafsson:2012vj,Blennow:2013pya,Law:2013saa,Hernandez:2013dta,Restrepo:2013aga,Chakraborty:2013gea,Ahriche:2014cda,Kanemura:2014rpa,Huang:2014bva,Varzielas:2015joa,Sanchez-Vega:2015qva,Fraser:2015mhb,Adhikari:2015woo,Biswas:2016bfo,Ahriche:2016rgf,Sierra:2016qfa,Lu:2016ucn,Batell:2016zod,Ho:2016aye,Escudero:2016ksa,Bonilla:2016diq,Borah:2016zbd,Biswas:2016yan,Hierro:2016nwm,Bhattacharya:2016qsg,Biswas:2016iyh,Chakraborty:2017dfg,Bhattacharya:2017sml,Ho:2017fte,Ghosh:2017fmr,Nanda:2017bmi,Narendra:2017uxl,Bernal:2017xat,Borah:2018gjk,Batell:2017cmf,Falkowski:2009yz,Falkowski:2011xh,Cherry:2014xra,Bertoni:2014mva,Allahverdi:2016fvl,Karam:2015jta,Bhattacharya:2018ljs,Biswas:2018sib,Gehrlein:2019iwl,Hashiba:2019mzm,Dasgupta:2019lha,Samanta:2020gdw}. One of the interesting possibilities is to connect the dark sector and the Standard Model through the RH neutrinos that realise the type I seesaw, which is usually named the neutrino portal scenario~\cite{Chianese:2018dsz,Chianese:2019epo,Chianese:2020khl,Becker:2018rve,Bian:2018mkl,Bandyopadhyay:2018qcv,Liu:2020mxj,Cosme:2020mck,Du:2020avz,Bandyopadhyay:2020qpn,Cheng:2020gut}. Some recent research~\cite{Chianese:2018dsz,Chianese:2019epo,Chianese:2020khl} shows that dark matter particles can be dominantly produced through the neutrino Yukawa interactions in the seesaw sector non-thermally by the so-called "freeze-in" mechanism~\cite{McDonald:2001vt,Hall:2009bx}. In those studies, the classical type I seesaw model is adopted and the right-handed neutrinos are superheavy in order to realise the leptogenesis~\cite{King:2018fqh}, which makes such a model hard to be constrained and tested by the relevant experiments~\cite{Drewes:2013gca,Drewes:2015vma,Drewes:2016jae,Deppisch:2015qwa,Chianese:2018agp,Beacham:2019nyx,SHiP:2018xqw,Blondel:2014bra}. Moreover, within the framework of the traditional type I seesaw model, even if the leptogenesis is not considered, the right-handed neutrinos are still required to be superheavy, otherwise the seesaw Yukawa coupling is too small to play a non-negligible role in dark matter production and the connection between neutrino physics and dark matter is lost. This motivates studies of DM in the type Ib seesaw model where the minimal 2RHN sector allows the simplest possible portal couplings, since this case has not been considered so far in the literature.

In this paper, we consider a minimal version of the type Ib seesaw model with 2RHNs which form a single heavy Dirac neutrino, 
within a type II 2HDM, where all fields transform under a $Z_3$ symmetry in such as way as to require two different Higgs doublets in the seesaw mechanism as shown in Fig.\ref{fig:Ib}. 
\footnote{
In the original type Ib seesaw model~\cite{Hernandez-Garcia:2019uof}, a $U(1)'$ symmetry controlled the neutrino sector, and with this symmetry
all Yukawa couplings were forbidden. However, in the present proposal, the $U(1)'$ is replaced by a $Z_3$ symmetry,
and the standard model fermions transform under $Z_3$ in such as way as to allow their renormalisable couplings to the two Higgs doublets.} 
We then discuss a simple dark matter extension of this model, in which the Dirac heavy neutrino is coupled to a dark Dirac fermion and a dark complex scalar field, both odd under a discrete $Z_2$ symmetry, where the lighter of the two is a dark matter candidate. 
To reduce the number of free parameters, we derive analytical formulae which show that the dark matter production does not depend on the individual mass of the dark scalar or dark fermion, but depends on the ratio of them, which agrees with the numerical result in previous works~\cite{Chianese:2018dsz,Chianese:2019epo}. 
Focussing on the fermionic dark matter case, and considering the freeze-in production of dark matter, we investigate the parameter space of type Ib seesaw Yukawa couplings, neutrino portal couplings and the ratio of dark particle masses which give the correct dark matter relic abundance. By considering the mixing between the standard model neutrinos and the heavy neutrino, we build a connection between the dark matter production and current laboratory experiments ranging from collider to lepton flavour violating experiments. 
For a GeV scale heavy neutrino, the parameters related to dark matter production are constrained by the experimental results directly and can be further tested by future experiments such as SHiP.

The paper is organised as follows. In Sec.\ref{sec:Model}, we briefly introduce the model studied in this paper and discuss its property and possible experimental constraints. In Sec.\ref{sec:DMpro}, we derive the Boltzmann equations and provides some analytical solutions. In Sec.\ref{sec:Results}, we present the numerical results from dark matter production and compare them to the existing experimental constraints and future experimental sensitivities. Finally, we summarise and conclude in Sec.\ref{sec:Concl}.

\section{Minimal type Ib seesaw model with dark matter \label{sec:Model}}

\begin{table}[t!]
\centering
\begin{tabular}{|c|c|c|c|c|c|c|c|c|c|c|c|}
\hline 
& ${Q}_\alpha$ & ${u_R}_\beta$ & ${d_R}_\beta$
& ${L}_\alpha$ & ${e_R}_\beta$ & $\Phi_{1}$ & $\Phi_{2}$  & $N_{\mathrm{R}1}$ & $N_{\mathrm{R}2}$ & $\phi$ & $\chi_{L,R}$ \\ \hline \hline 
$SU(2)_L$ & {\bf 2} & {\bf 1} & {\bf 1} & {\bf 2} & {\bf 1} & {\bf 2} & {\bf 2} & {\bf 1} & {\bf 1} & {\bf 1} & {\bf 1} \\ \hline
$U(1)_Y$ & $\frac{1}{6}$ & $\frac{2}{3}$ & $-\frac{1}{3}$ & $-\frac{1}{2}$ & $-1$ & $-\frac{1}{2}$ & $-\frac{1}{2}$ & 0 & 0 & 0 & 0 \\ \hline  \hline
$Z_3$ & $1$ &  $\omega$ &   $\omega$ & $1$ &  $\omega$ &   $\omega$ &   $\omega^2$ & $\omega^2$ &   $\omega$ &  $\omega$  & $\omega^2$  \\ \hline  
$Z_2$ & $+$ & $+$ & $+$ & $+$ & $+$ & $+$ & $+$ & $+$ & $+$ & $-$ & $-$ \\ \hline
\end{tabular}
\caption{\label{tab:matter}Irreducible representations of the fields of the model under the electroweak $SU(2)_L\times U(1)_Y$ gauge symmetry, the discrete $Z_3$ symmetry (where we write $\omega =e^{i2\pi/3}$) and the unbroken $Z_2$
dark symmetry. The fields $Q_{\alpha}, L_{\alpha}$ are left-handed SM doublets while ${u_R}_\beta,{d_R}_\beta,{e_R}_\beta$
are right-handed SM singlets where $\alpha, \beta = 1,2,3$ label the three families of quarks and leptons. The fields $N_{\mathrm{R}{1,2}}$ are the two right-handed neutrinos, while $\phi$ and $\chi_{L,R}$ are a dark complex scalar and dark Dirac fermion, respectively.}
\end{table}

Here we introduce the minimal version of the type Ib seesaw model with 2RHNs~\cite{Hernandez-Garcia:2019uof}, 
where all fields transform under a $Z_3$ symmetry in such as way as to require two different Higgs doublets in the seesaw mechanism, as shown in Fig.\ref{fig:Ib}. 
The two right-handed neutrinos $N_{\mathrm{R}1}$ and $N_{\mathrm{R}2}$ form a single heavy Dirac neutrino $\N$. We also consider a dark matter extension of this model to include a $Z_2$-odd dark sector containing a singlet Dirac fermion $\chi$ and a singlet complex scalar $\phi$. 
The fields of the model are summarized in Tab.~\ref{tab:matter}. The $Z_3$ symmetry ensures that the coupling between the Higgs doublets and SM fermions follows the type II 2HDM pattern: The masses of the charged leptons and $-1/3$ charged quarks are generated by the spontaneous symmetry breaking (SSB) of the first Higgs doublet $\Phi_1$, while the $2/3$ charged quarks gain masses from $\Phi_2$. The full Lagrangian can be separated into parts as
\begin{equation}
\mathcal{L} = \mathcal{L}_{\rm 2HDM}  + \mathcal{L}_{\rm seesaw Ib} + \mathcal{L}_{\rm DS} + \mathcal{L}_{\rm N_Rportal} \,.
\label{eq:lag}
\end{equation}
The first term is the Lagrangian of the 2HDM which includes couplings between charged fermions and Higgs doublets  
\begin{eqnarray}
\mathcal{L}_{\rm 2HDM} & \supset & - Y^u_{\alpha \beta} \overline{Q}_\alpha { \Phi}_2 u_{R\beta}
- Y^d_{\alpha \beta} \overline{Q}_\alpha {\tilde \Phi}_1 d_{R\beta}- Y^e_{\alpha \beta} \overline{L}_\alpha {\tilde \Phi}_1 e_{R\beta}
 + {\rm h.c.}\,
\label{eq:Yuk} 
\end{eqnarray}
The remaining terms are the type Ib seesaw Lagrangian, dark sector (DS) Lagrangian and the neutrino portal, which take the form 
\begin{eqnarray}
\mathcal{L}_{\rm seesawIb} & = & - Y_{1\alpha} \overline{L}_\alpha {\Phi}_1 N_{R1} 
- Y_{2\alpha} \overline{L}_\alpha {\Phi}_2 N_{R2}
- M_N\overline{N^c_{R1}} N_{R2} + {\rm h.c.}\,, \\
\mathcal{L}_{\rm DS} & = & \overline{\chi}\left(i \slashed{\partial} - m_\chi \right)\chi + \left|\partial_\mu \phi\right|^2 - m^2_\phi \left|\phi\right|^2 + V\left(\phi\right)\,, 
\label{eq:lagDS} \\
\mathcal{L}_{\rm N_R portal} & = &  y_{1} \phi \, \overline{\chi_R}N^c_{R1} + y_{2} \phi \, \overline{\chi_L}N_{R2} + {\rm h.c.} \,, 
\label{eq:ds}
\end{eqnarray}
The two ``right-handed'' Weyl neutrinos can actually form a four component Dirac spinor $\N = \left( N^c_{R1},N_{R2} \right)$ with a Dirac mass $M_N$. Moreover, if the two portal couplings are equal $y = y_1 = y_2$, the neutrino portal has a CP invariant form after $\N$ is introduced. The type Ib seesaw Lagrangian and the neutrino portal can be rewritten as 
\begin{eqnarray}
\mathcal{L}_{\rm seesawIb} & = & - Y_{1\alpha}^* \overline{L^c}_\alpha \Phi_1^* \N_L - Y_{2\alpha} \overline{L}_\alpha {\Phi}_2 \N_R - M_N\overline{\N_L} \N_R+ {\rm h.c.}\,, 
\label{eq:lagNS}\\
\mathcal{L}^{\rm Parity}_{\rm N_R portal} & = &  y \phi \, \overline{\chi}\N   + {\rm h.c.} \,,
\label{eq:lagPortal}
\label{eq:ds}
\end{eqnarray}
where $\chi=\left( \chi_L,\chi_R \right)$. Eq.\eqref{eq:lagDS} defines the kinetic and mass terms of the dark particles as well as a general potential of the dark scalar $\phi$. The vacuum is required to appear at zero in the potential so that the $Z_2$ symmetry is preserved. The neutrino portal in Eq.\eqref{eq:lagPortal} includes a Yukawa-like coupling between the heavy neutrino in the visible sector and dark particles. Although the couplings between the Higgs doublets and the dark scalar is also not forbidden by the discrete symmetries, we assume those couplings are negligible\footnote{The Higgs portal couplings and constraints are discussed 
in the Appendix.} and in this study and we focus on the neutrino portal.

In the type Ib seesaw Lagrangian, the heavy neutrino can be integrated out to generate a set of effective operators~\cite{Broncano:2002rw}, which leads to an effective field theory for the low energy phenomenology. The dimension five effective operators are Weinberg-type operators involving two different Higgs doublets~\cite{Hernandez-Garcia:2019uof}
\begin{eqnarray}
\delta \mathcal{L}^{d=5} &=& 
c^{d=5}_{\alpha \beta}
\left( (\overline{L^c}_\alpha \Phi_1^*) (\Phi_2^\dagger {L}_\beta ) + (\overline{L^c}_\beta \Phi_1^*) (\Phi_2^\dagger {L}_\alpha )\right)
\nonumber\\&&+ \left(c^{d=5}_{\alpha \beta}\right)^*
\left(\left(\overline{L}_\beta \Phi_2 \right)\left(\Phi_1^T {L^c}_\alpha\right) + \left(\overline{L}_\alpha \Phi_2 \right)\left(\Phi_1^T {L^c}_\beta\right) \right)\,.
\label{eq:d_5op}
\end{eqnarray}
Different from the type Ia seesaw model, the standard Weinberg operator with two $\Phi_1$ or two $\Phi_2$ is forbidden by the $Z_3$ symmetry and that only the new Weinberg-type operator that mixes the two Higgs doublets is allowed in the model. When the Higgs doublets develop VEVs as $\langle\Phi_i\rangle=\left(v_i/\sqrt{2},0\right) $, the new Weinberg-type operator induces Majorana mass terms $m_{\alpha \beta} \nu_{\alpha} \nu_{\beta}$ for the light SM neutrinos, where 
\begin{equation}
m_{\alpha \beta}= v_1  v_2 c^{d=5}_{\alpha \beta}= \dfrac{  v_1  v_2}{M_N}\left(Y_{1\alpha}^* Y_{2\beta}^*  + Y_{1\beta}^* Y_{2\alpha}^* \right)\,,
\label{eq:dim5_relation_simp}
\end{equation} 
The smallness of the light neutrino masses may stem not only from the suppression of $M_N$, but also from the Yukawa couplings. Since there are two different Yukawa couplings, one of them can be sizeable if the other one is small enough, allowing a low scale seesaw model. This enables the seesaw mechanism to play a role in dark matter even for GeV mass heavy neutrinos.

It has been shown that, similar to the simplest minimal flavor violating type-I seesaw model~\cite{Gavela:2009cd}, the Yukawa couplings $Y_{1\alpha}, Y_{2\alpha}$ in type Ib seesaw model can be determined by the elements of the PMNS mixing matrix $U_\text{PMNS}$ the two mass squared splittings $\Delta m_\text{sol}^2$ and $\Delta m_\text{atm}^2$, up to overall factors $Y_1$ and $Y_2$~\cite{Hernandez-Garcia:2019uof}. In this minimal scenario, only two neutrinos get masses and the lightest neutrino remains massless. On the other hand, considering the hierarchy of the neutrinos is undetermined, there are two distinguishable possibilities for the Yukawa couplings. For the case of a normal hierarchy (NH), the Yukawa couplings in the flavour basis, where the charged lepton mass matrix is diagonal, read
\begin{eqnarray}
Y_{1\alpha} &=&\dfrac{Y_1}{\sqrt{2}}\left( \sqrt{1+\rho} \left(U_\text{PMNS}\right)_{\alpha 3}-\sqrt{1-\rho} \left(U_\text{PMNS}\right)_{\alpha 2}\right)\,, 
\label{eq:Y1_NH}\\
Y_{2\alpha} &=& \dfrac{Y_2}{\sqrt{2}}\left( \sqrt{1+\rho} \left(U_\text{PMNS}\right)_{\alpha 3}+\sqrt{1-\rho} \left(U_\text{PMNS}\right)_{\alpha 2}\right)\,, 
\label{eq:Y2_NH}
\end{eqnarray}
where $Y_2,\,Y_1$ are real numbers and  $\rho={(\sqrt{1+r}-\sqrt{r})}/{(\sqrt{1+r}+\sqrt{r})}$ with $r \equiv {\vert\Delta m_{21}^2\vert}/{\vert\Delta m_{32}^2\vert}$.
The neutrino masses in the NH are
\begin{eqnarray}
m_1=0, \ \ \vert m_2 \vert=\frac{Y_1Y_2v_1v_2}{M_N}(1-\rho),\ \ \vert m_3\vert=\frac{Y_1Y_2v_1v_2}{M_N}(1+\rho).
\end{eqnarray}
For an inverted hierarchy (IH), the Yukawa couplings in the flavour basis are given by
\begin{eqnarray}
Y_{1\alpha}&=&\dfrac{Y_1}{\sqrt{2}}\left( \sqrt{1+\rho} \left(U_\text{PMNS}\right)_{\alpha2}-\sqrt{1-\rho} \left(U_\text{PMNS}\right)_{\alpha1}\right)\,, \label{eq:Yukawa_relation_IH} \\
Y_{2\alpha}&=&\dfrac{Y_2}{\sqrt{2}}\left( \sqrt{1+\rho} \left(U_\text{PMNS}\right)_{\alpha 2}+\sqrt{1-\rho} \left(U_\text{PMNS}\right)_{\alpha 1}\right)\,, 
\end{eqnarray}
where $\rho={(\sqrt{1+r}-1)}/{(\sqrt{1+r}+1)}$ with $r\equiv {\vert\Delta m_{21}^2\vert}/{\vert \Delta m_{31}^2 \vert}$. The neutrino masses in the IH are
\begin{eqnarray}
m_3=0, \ \ \vert m_1 \vert =\frac{Y_1Y_2v_1v_2}{M_N}(1-\rho),\ \ \vert m_2 \vert =\frac{Y_1Y_2v_1v_2}{M_N}(1+\rho).
\end{eqnarray}
Since only the overall factors $Y_1$ and $Y_2$ are unfixed, we refer to $Y_1$ and $Y_2$ as Yukawa couplings in the rest of the paper. With the central values of oscillation parameters~\cite{Esteban:2020cvm} and setting $\delta_{\rm CP}=\pi$, a combined value is fixed as
\begin{eqnarray}
\frac{Y_1Y_2v_1v_2}{M_N} = 
\begin{dcases} 
3.0 \times 10^{-11}\, \text{GeV} & \text{for NH} \\ 
5.0 \times 10^{-11}\, \text{GeV} & \text{for IH} 
\end{dcases}.
\label{eq:rel1}
\end{eqnarray}
In 2HDM, it is common to define the ratio of Higgs VEVs as $\tan \beta =v_2/v_1$, where $v_1$ and $v_2$ are the VEVs of $\Phi_1$ and $\Phi_2$ respectively. Assuming there is no complex relative phase between the VEVs, the Higgs VEVs follow the relation $\sqrt{v_1^2+v_2^2} = v = 246$ GeV and Eq.\eqref{eq:rel1} can be simplified 
\begin{eqnarray}
\frac{Y_1 Y_2 \sin{2\beta}}{M_N} = 
\begin{dcases} 1.0 \times 10^{-15}\, \text{GeV}^{-1} & \text{for NH} \\ 
1.6 \times 10^{-15}\, \text{GeV}^{-1} & \text{for IH} 
\end{dcases}.
\label{eq:rel2}
\end{eqnarray}
In summary, there are four free parameters constrained by one relation Eq.\eqref{eq:rel2} in the minimal type Ib seesaw model. As will be shown later, the quantity that matters in dark matter production is the sum of the squared Yukawa couplings instead of their product. Therefore it is useful to derive a lower limit to the sum of squared Yukawa couplings from Eq.\eqref{eq:rel2} using the inequality of arithmetic and geometric means (AM–GM inequality)
\begin{eqnarray}
Y_1^2 + Y_2^2  \gtrsim
\begin{dcases} 2.0 \times 10^{-15}\, \frac{M_N}{1 \text{GeV}} \frac{1}{\sin{2\beta}}& \text{for NH} \\ 
3.3 \times 10^{-15}\, \frac{M_N}{1 \text{GeV}} \frac{1}{\sin{2\beta}} & \text{for IH} 
\end{dcases}.
\label{eq:rel3}
\end{eqnarray}
For simplicity, we focus on the normal hierarchy of neutrino mass ordering from now on and discuss the experimental constraints on the model.

\subsection{$\mu \rightarrow \gamma e$}
The first experimental constraint on type Ib seesaw model is from the $\mu \rightarrow \gamma e$ decay.\footnote{The strongest constraint from charged lepton decay is the one from the muon decay. In the type Ib seesaw model, the quantities $\left| \eta_{e\mu} \right|$ and $\left| \eta_{\mu\tau} \right|$ have the same order of magnitude ($\left| \eta_{e\mu} \right|/\left| \eta_{\mu\tau} \right| \sim 0.5$), while experimental constraint from $\mu$ decay on $\left| \eta_{e\mu} \right|$ is $10^{-3}$ smaller than the one from $\tau$ decay on $\left| \eta_{\mu\tau} \right|$ \cite{Fernandez-Martinez:2016lgt}.} In the framework of type Ib seesaw model, the total $5\times 5$ mass matrix of neutrinos is given 
in the flavour basis, where the charged lepton mass matrix is diagonal,
by~\cite{Hernandez-Garcia:2019uof} 
\begin{eqnarray}
&&\begin{matrix}\phantom{Y_{11}}\nu_e & \phantom{Y_{11}}\nu_\mu & \phantom{Y_{11}}\nu_\tau & \phantom{Y_{11}}\N_L & \phantom{Y_{1}}\N_R^c \end{matrix} \nonumber\\
M^\nu = 
\begin{matrix} \overline{\nu^c_e} \\[10pt] \overline{\nu^c_\mu} \\[10pt] \overline{\nu^c_\tau} \\[10pt] \overline{\N_L^c} \\[10pt] \overline{\N_R} \end{matrix}
&&\begin{pmatrix} 
0 & 0 & 0 &\dfrac{Y_{11}^* v_1}{\sqrt{2}} & \dfrac{Y_{21}^* v_2}{\sqrt{2}} \\[8pt] 
0 & 0 & 0 &\dfrac{Y_{12}^* v_1}{\sqrt{2}} & \dfrac{Y_{22}^* v_2}{\sqrt{2}} \\[8pt] 
0 & 0 & 0 &\dfrac{Y_{13}^* v_1}{\sqrt{2}} & \dfrac{Y_{23}^* v_2}{\sqrt{2}} \\[8pt]
\dfrac{Y_{11}^* v_1}{\sqrt{2}} & \dfrac{Y_{12}^* v_1}{\sqrt{2}} & \dfrac{Y_{13}^* v_1}{\sqrt{2}} &0 & M_N \\[8pt] 
\dfrac{Y_{21}^* v_2}{\sqrt{2}} & \dfrac{Y_{22}^* v_2}{\sqrt{2}} & \dfrac{Y_{23}^* v_2}{\sqrt{2}} & M_N & 0 \end{pmatrix}\equiv \begin{pmatrix} 0 & m_D^T \\ m_D & M \end{pmatrix}\, .
\label{Mnu}
\end{eqnarray}
$M^\nu$ is a symmetric complex matrix and thus can be diagonalised by a unitary transformation $U$ of the form $U^TM^\nu U$. 
In the flavour basis, the $5\times 5$ unitary matrix $U$ takes the approximate expression~\cite{Hernandez-Garcia:2019uof}
\begin{eqnarray}
U &\simeq & 
\begin{pmatrix} I_{3\times 3}-\dfrac{\Theta\Theta^\dagger}{2} & \Theta \\ -\Theta^\dagger & I_{2\times 2} - \dfrac{\Theta^\dagger\Theta}{2} \end{pmatrix}
\begin{pmatrix} U_\text{PMNS} & 0 \\ 0 & I_{2\times 2} \end{pmatrix}
\label{eq:matrixU}
\end{eqnarray}
where $\Theta = m_D^\dagger M^{-1}$ is a $3\times2$ matrix 
in this model, while and $I_{3\times 3}$, $I_{2\times 2}$ are unit matrices of the specified dimension. The $e\mu$ element of the hermitian matrix $\eta$ defined by $\eta=\Theta\Theta^\dagger/2$ is constrained by $\mu \rightarrow \gamma e$ through the neutrino mixing~\cite{Hernandez-Garcia:2019uof}
\begin{eqnarray}
\left| \eta_{e\mu} \right| = \frac{\left| Y_{1e} Y_{1\mu}^* v_1^2 + Y_{2e} Y_{2\mu}^* v_2^2 \right|} {4 M_N^2} \lesssim 8.4 \times 10^{-6}\,.
\label{eq:mudecay}
\end{eqnarray}
Besides $Y_1$ and $Y_2$, it is clear from Eq.\eqref{eq:Y1_NH} and Eq.\eqref{eq:Y2_NH} that the mixing between $\nu_e$ and $\nu_\mu$ also depends on the unconstrained relative Majorana phase $\delta_M$ in the PMNS mixing matrix. 

\subsection{ Neutrino mixing}

For sub-TeV heavy neutrino masses $M_N$, the mixing between the SM neutrinos and the heavy neutrino is also constrained by existing collider data~\cite{Drewes:2013gca,Drewes:2015vma,Drewes:2016jae} as well as future experiments~\cite{Deppisch:2015qwa,Chianese:2018agp,Beacham:2019nyx} like the SHiP experiment~\cite{SHiP:2018xqw} and FCC-$ee$~\cite{Blondel:2014bra}. The strength of the mixing between SM neutrinos and the heavy neutrino is represented by the quantity 
\begin{eqnarray}
U_\alpha^2=\sum_{i=L,R}|U_{\alpha i}|^2\,, \quad \alpha=e,\mu,\tau,
\label{eq1:U2}
\end{eqnarray}
where $U$ is the $5\times 5$ unitary matrix in Eq.\eqref{eq:matrixU} and in the above expression we have 
summed over the two heavy neutrino indices ${\N_L^c}$, ${\N_R}$ for each light neutrino flavour 
$\nu_e,\nu_{\mu},\nu_{\tau}$. 
More specifically, for the SM neutrinos $\nu_e,\nu_{\mu},\nu_{\tau}$ in the flavour basis Eqs. \ref{eq:Y1_NH}, \ref{eq:Y2_NH}, \ref{Mnu}, \ref{eq:matrixU}, \ref{eq1:U2} give for central values of oscillation parameters~\cite{Esteban:2020cvm}
\begin{subequations}
\label{eq2:U2}
\begin{align}
U_e^2&= \frac{ (0.031+0.029\cos\delta_M) v_1^2 Y_1^2 + (0.031-0.029\cos\delta_M) v_2^2 Y_2^2}{M_N^2}\,,\\
U_\mu^2&= \frac{ (0.27-0.16\cos\delta_M) v_1^2 Y_1^2 + (0.27+0.16\cos\delta_M) v_2^2 Y_2^2}{M_N^2}\,,\\
U_\tau^2&= \frac{ (0.20+0.13\cos\delta_M) v_1^2 Y_1^2 + (0.20-0.13\cos\delta_M) v_2^2 Y_2^2}{M_N^2}\,,
\end{align}
\end{subequations}
where $\delta_M$ is the unmeasured relative Majorana phase. If only one of the seesaw Yukawa couplings dominates, the quantity $U^2_\alpha$ is proportional to $v_iY_i/M_N$, where $i=1,\,2$ depending on which Yukawa coupling is dominating. Using Eq.\eqref{eq:rel1}, the dependence on one of the Yukawa couplings can be removed and lower limits of $U_\alpha^2$ can be obtained. For example, removing $Y_1$ leads to simplification of Eq.\eqref{eq2:U2} as
\begin{subequations}
\label{eq3:U2}
\begin{align}
U_e^2&=(3.0 \times 10^{-11})^2 (0.031+0.029\cos\delta_M) \left(\frac{1 \text{GeV}}{v_2 Y_2}\right)^{2} +  (0.031-0.029\cos\delta_M) \left(\frac{v_2 Y_2}{M_N}\right)^2 \nonumber
\\ &\geq 5.9\times 10^{-13} \sqrt{9.6-8.2\cos^2\delta_M}\, \frac{1 \text{GeV}}{M_N}\,,\\
U_\mu^2&=(3.0 \times 10^{-11})^2 (0.27-0.16\cos\delta_M) \left(\frac{1 \text{GeV}}{v_2 Y_2}\right)^{2}  + (0.27+0.16\cos\delta_M) \left(\frac{v_2 Y_2}{M_N}\right)^2  \nonumber
\\ &\geq 5.9\times 10^{-12} \sqrt{7.0-2.6\cos^2\delta_M}\, \frac{1 \text{GeV}}{M_N}\,,\\
U_\tau^2&=(3.0 \times 10^{-11})^2 (0.20+0.13\cos\delta_M) \left(\frac{1 \text{GeV}}{v_2 Y_2}\right)^{2}  + (0.20-0.13\cos\delta_M) \left(\frac{v_2 Y_2}{M_N}\right)^2   \nonumber
\\ &\geq 5.9\times 10^{-12} \sqrt{4.2-1.7\cos^2\delta_M}\, \frac{1 \text{GeV}}{M_N}\,,
\end{align}
\end{subequations}
where the inequalities are the application of the AM–GM inequality. It can be deduced from Eq.\eqref{eq3:U2} that the lowest allowed value of $U^2_\alpha$ is achieved when $\cos^2\delta_M = 1$. For each neutrino flavour, the minimum is achieved when 
\begin{subequations}
\label{eq:U2minY}
\begin{align}
Y_2&=1.2\times10^{-5}\sqrt{M_N/v_2} \quad \text{or} \quad Y_1=2.4\times10^{-6}\sqrt{M_N/v_1} \quad \text{for $e$ neutrino,} \\
Y_2&=3.8\times10^{-6}\sqrt{M_N/v_2} \quad \text{or} \quad Y_1=7.7\times10^{-6}\sqrt{M_N/v_1} \quad \text{for $\mu$ neutrino,} \\
Y_2&=8.0\times10^{-6}\sqrt{M_N/v_2} \quad \text{or} \quad Y_1=3.7\times10^{-6}\sqrt{M_N/v_1} \quad \text{for $\tau$ neutrino,}
\end{align}
\end{subequations}
for the cases $v_1Y_1\ll v_2Y_2$ and $v_1Y_1\gg v_2Y_2$, respectively. The derivation of such a lower limit of $U^2_\alpha$ relies on the particular relation of the seesaw couplings Eq.\eqref{eq:rel1} in the framework of type Ib seesaw model and thus it is distinguishable from the minimum $U^2_\alpha$ required in other types of seesaw models.

\section{Dark matter production \label{sec:DMpro}}

In principle, both the dark scalar and dark fermion can be dark matter candidate, depending on their masses. For simplicity, we focus on the mass case where the dark scalar is heavier than the dark fermion and we require the dark scalar is heavy enough to decay into the dark fermion and heavy neutrinos, i.e. $m_\phi>m_\chi+M_N$, to keep a single DM scenario. In general, both the freeze-out and freeze-in mechanism can produce the correct dark matter relic density. In this work, we focus on the freeze-in and assume neglectable comoving number density of dark particles at the end of reheating.

\begin{figure}[t!]
\begin{center}
\subfigure[~Neutrino Yukawa processes]{
\begin{fmffile}{NS1}
\fmfframe(23,18)(18,18){
\begin{fmfgraph*}(80,45)
\fmflabel{$\phi^*$}{o1}
\fmflabel{$\chi$}{o2}
\fmflabel{$\nu_\alpha,\ell^+_\alpha$}{i2}
\fmflabel{$\phi_1^0,\phi_1^-$}{i1}
\fmfv{label=$y$}{v2}
\fmfv{label=$Y_{1\alpha}$}{v1}
\fmfleft{i1,i2}
\fmfright{o1,o2}
\fmf{plain}{i2,v1}
\fmf{dashes}{v2,o1}
\fmf{fermion}{v2,o2}
\fmf{dashes}{i1,v1}
\fmf{fermion,label=$\N$}{v1,v2}
\fmfdotn{v}{2}
\end{fmfgraph*}}
\end{fmffile}
\begin{fmffile}{NS2}
\fmfframe(23,18)(18,18){
\begin{fmfgraph*}(80,45)
\fmflabel{$\phi^*$}{o1}
\fmflabel{$\chi$}{o2}
\fmflabel{$\nu_\alpha,\ell^-_\alpha$}{i2}
\fmflabel{${\phi_2^0},\phi_2^+$}{i1}
\fmfv{label=$y$}{v2}
\fmfv{label=$Y_{2\alpha}^*$}{v1}
\fmfleft{i1,i2}
\fmfright{o1,o2}
\fmf{plain}{i2,v1}
\fmf{dashes}{v2,o1}
\fmf{fermion}{v2,o2}
\fmf{dashes}{i1,v1}
\fmf{fermion,label=$\N$}{v1,v2}
\fmfdotn{v}{2}
\end{fmfgraph*}}
\end{fmffile}\label{fig:feynA}}
\subfigure[~Dark sector processes independent of neutrino Yukawa couplings]{
\begin{fmffile}{DS1}
\fmfframe(20,20)(20,20){
\begin{fmfgraph*}(80,45)
\fmflabel{$\phi$}{o1}
\fmflabel{$\phi^*$}{o2}
\fmflabel{$\overline{\N}$}{i1}
\fmflabel{$\N$}{i2}
\fmfv{label=$y$}{v1}
\fmfv{label=$y$}{v2}
\fmfleft{i1,i2}
\fmfright{o1,o2}
\fmf{fermion}{i2,v1}
\fmf{dashes}{v1,o2}
\fmf{fermion}{v2,i1}
\fmf{dashes}{v2,o1}
\fmf{fermion,label=$\chi$,tension=0}{v1,v2}
\fmfdotn{v}{2}
\end{fmfgraph*}}
\end{fmffile}
\begin{fmffile}{DS2}
\fmfframe(20,20)(20,20){
\begin{fmfgraph*}(80,45)
\fmflabel{$\overline{\chi}$}{o1}
\fmflabel{$\chi$}{o2}
\fmflabel{$\overline{\N}$}{i1}
\fmflabel{$\N$}{i2}
\fmfv{label=$y$}{v1}
\fmfv{label=$y$}{v2}
\fmfleft{i1,i2}
\fmfright{o1,o2}
\fmf{fermion}{i2,v1}
\fmf{fermion}{v1,o2}
\fmf{fermion}{v2,i1}
\fmf{fermion}{o1,v2}
\fmf{dashes,label=$\phi$,tension=0}{v1,v2}
\fmfdotn{v}{2}
\end{fmfgraph*}}
\end{fmffile}
\label{fig:feynB}}
\end{center}
\caption{\label{fig:Feyn} 
Processes responsible for the dark matter production considered in this study. 
}
\end{figure}
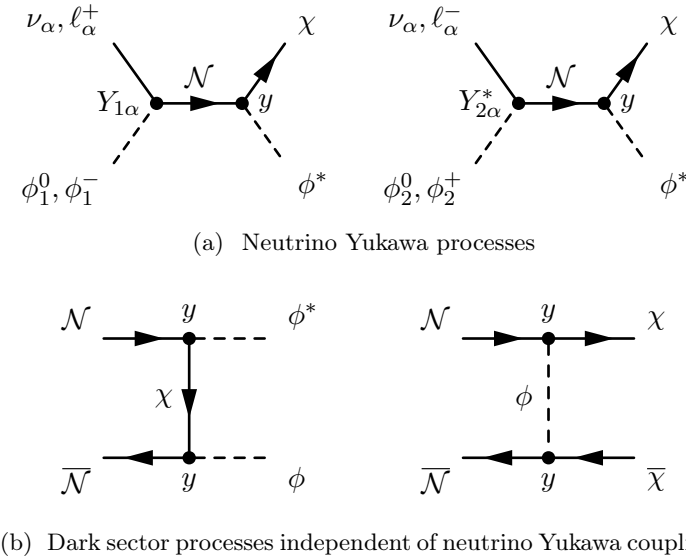

The Feynman diagrams for processes that are relevant to dark matter production are shown in Fig.\ref{fig:Feyn}. There are two classes of processes named as neutrino Yukawa processes and dark sector processes, respectively. The neutrino Yukawa processes are the scattering between SM particles into one dark scalar and one dark fermion, mediated by the heavy neutrino, while the dark sector processes are the scattering of two heavy neutrinos into two dark scalars or two dark fermions. 

The evolution of the dark particle number density follows the Boltzmann equation. Here we use a variation of the Boltzmann equation which shows the evolution of yield $Y$ as a function of the photon temperature $T$. The yield $Y$ is defined as the ratio of the number density and the entropy density, $Y\equiv n/\mathfrak{s}$. The Boltzmann equations for the dark particles are given by
\begin{eqnarray}
\mathcal{H}\,T\left(1+\frac{T}{3 g^\mathfrak{s}_*\left(T\right)}\frac{d g^\mathfrak{s}_*}{d T}\right)^{-1} \frac{d Y_\phi}{d T} & = & 
- \mathfrak{s} \left<\sigma\, v\right>_{\phi\phi}^{\rm DS} \left({Y_\phi^{\rm eq}}\right)^2 
- \mathfrak{s} \left<\sigma\, v\right>^{\rm \nu-Yukawa}_{\chi\phi} Y_\phi^{\rm eq}Y_\chi^{\rm eq} \nonumber\\&& + \left<\Gamma_{\phi}\right>\left(Y_\phi-\frac{Y_\phi^{\rm eq}}{Y_\chi^{\rm eq}}Y_\chi\right)  \,, \label{eq:Boltphi} \\
\mathcal{H}\,T\left(1+\frac{T}{3 g^\mathfrak{s}_*\left(T\right)}\frac{d g^\mathfrak{s}_*}{d T}\right)^{-1} \frac{d Y_\chi}{d T} & = & 
- \mathfrak{s}  \left<\sigma\, v\right>_{\chi\chi}^{\rm DS} \left({Y_\chi^{\rm eq}}\right)^2 
- \mathfrak{s} \left<\sigma\, v\right>^{\rm \nu-Yukawa}_{\chi\phi} Y_\phi^{\rm eq}Y_\chi^{\rm eq} \nonumber\\&&  -  \left<\Gamma_{\phi}\right>\left(Y_\phi-\frac{Y_\phi^{\rm eq}}{Y_\chi^{\rm eq}}Y_\chi\right) \,, \label{eq:Boltchi}
\end{eqnarray}
where $\left<\sigma\, v\right>$ is the thermal averaged cross section and $\left<\Gamma\right>$ is the thermal averaged decay rate. The superscripts ``DS'' and ``$\nu$-Yukawa'' refer to the total contributions from the dark sector processes and the neutrino Yukawa ones, respectively. The heavy neutrino $\N$ is assumed to be in thermal equilibrium. The Boltzmann equation for total dark matter yield is obtained by adding Eq.\eqref{eq:Boltphi} and Eq.\eqref{eq:Boltchi} together
 \begin{eqnarray}
\mathcal{H}\,T\left(1+\frac{T}{3 g^\mathfrak{s}_*\left(T\right)}\frac{d g^\mathfrak{s}_*}{d T}\right)^{-1} \frac{d Y_{\rm DM}}{d T} & = & 
- \mathfrak{s} \left<\sigma\, v\right>_{\phi\phi}^{\rm DS} \left({Y_\phi^{\rm eq}}\right)^2 - \mathfrak{s}  \left<\sigma\, v\right>_{\chi\chi}^{\rm DS} \left({Y_\chi^{\rm eq}}\right)^2 
\nonumber\\&& -2 \mathfrak{s} \left<\sigma\, v\right>^{\rm \nu-Yukawa}_{\chi\phi} Y_\phi^{\rm eq}Y_\chi^{\rm eq}\,.  \label{eq:DM}
\end{eqnarray}
The contribution from dark scalar decay is cancelled as it does not change the number of dark matter particles. The final yield should meet the yield of dark matter today, which can be calculated with the observed relic abundance $\Omega_{\rm DM}h^2$, entropy density $\mathfrak{s}_0$ and critical density $\rho_{\rm crit}/h^2$
\begin{equation}
Y_{\rm DM,0} = \Omega_{\rm DM}h^2 \frac{\rho_{\rm crit}/h^2}{2 \, \mathfrak{s}_0 \, m_\chi }  \,.
\label{eq:omegaPREA}
\end{equation}
The observed relic abundance is provided by the Planck Collaboration at 68\% C.L.~\cite{Aghanim:2018eyx}:
\begin{equation}
\Omega_{\rm DM}^\mathrm{obs}h^2 = 0.120 \pm 0.001\,.
\label{eq:omegaOBS}
\end{equation}

In general, the Boltzmann equation can be solved numerically. However, we would like to derive some analytical results which can confirm and be confirmed by the numerical results later. In the limit $m_\phi \gg M_N\,,m_\chi$, the scattering amplitude for the contribution from $\nu$-Yukawa processes is 
\begin{eqnarray}
\frac12\sum_{\text{internal d.o.f.}}\int \overline{\left|\mathcal{M}\right|^2}_{\nu-{\rm Yukawa}} d\Omega & = & 
2 \pi y^2 \left( Y_1^2 + Y_2^2 \right) \left(1 - \frac{ m_\phi^2}{s}\right)\,. 
\end{eqnarray}
Then, with the approximation of Maxwell-Boltzmann distribution for all the particles,\footnote{ As shown in \cite{Lebedev:2019ton}, such an approximation may cause up to around 50\%  difference in the reaction rates for relativistic particles.} the thermal averaged cross section is
\begin{eqnarray}
\left<\sigma\, v\right>^{\rm \nu-Yukawa}_{\chi\phi} &=& \frac{g_\chi g_\phi}{n^{\rm eq}_{\chi}\, n^{\rm eq}_{\phi}}\frac{y^2 \left( Y_1^2 + Y_2^2 \right) T}{1024\pi^5} 
\int_{m_\phi^2}^\infty ds \, \sqrt{s-m_\phi^2}  \,K_1\left(\sqrt{s}/T\right) \left(1 - \frac{m_\phi^2}{s}\right) \,\nonumber\\
&=&  \frac{g_\chi g_\phi}{n^{\rm eq}_{\chi}\, n^{\rm eq}_{\phi}}\frac{y^2 \left( Y_1^2 + Y_2^2 \right) T^4}{1024\pi^5} \times \nonumber\\&&
\int_{(m_\phi/T)^2}^\infty ds' \, \sqrt{s'-(m_\phi/T)^2}  \,K_1\left(\sqrt{s'}\right) \left(1 - \frac{(m_\phi/T)^2}{s'}\right) \,, 
\label{eq:avgcsYuk1}
\end{eqnarray}
where the second step is rescaling the variable from $s \rightarrow T^2s'$. The new variable is $s'$ is dimensionless as well as the integral. Therefore the integral only depends on the dimensionless quantity $m_\phi/T$ and can be replaced by a function $F(T/m_\phi)$ defined as the antiderivative of the integrand
\begin{eqnarray}
\left<\sigma\, v\right>^{\rm \nu-Yukawa}_{\chi\phi} 
&=& \frac{g_\chi g_\phi}{n^{\rm eq}_{\chi}\, n^{\rm eq}_{\phi}}\frac{ y^2 \left( Y_1^2 + Y_2^2 \right) T^4}{1024\pi^5} F(T/m_\phi)\,.
\end{eqnarray}
Then the yield of DM through $\nu$-Yukawa processes is
\begin{eqnarray}
Y_{\rm DM}^{\nu-\rm Yukawa} &=& \frac{y^2 \left( Y_1^2 + Y_2^2 \right)}{256\pi^5}\int_0^{T_{\rm RH}} dT \frac{T^3}{\mathcal{H}\,\mathfrak{s}}  F(T/m_\phi)
= \frac{y^2 \left( Y_1^2 + Y_2^2 \right)}{256\pi^5 m_\phi} C \int_0^{T_{\rm RH}/m_\phi} dT' \frac{ F(T')}{T'^2}\,,\nonumber\\&& 
\end{eqnarray}
where $C = T^5/(H \mathfrak{s}) \simeq 1.5 \times 10^{16}$ GeV is a constant. In the second step, the variable $T$ is rescaled as $T \rightarrow m_\phi T'$. Again, $T'$ is dimensionless and the integral only depends on its upper limit $T_{\rm RH}/m_\phi$. If the reheating temperature is high enough ($T_{\rm RH} \gg m_\phi$), the integral above is not sensitive to the upper limit and behaves like a constant. Notice that the change in relativistic degrees of freedom caused by the decouple of dark particles is neglected since it only contributes few percents to the final results. After all these approximations are made, the yield of dark matter through $\nu$-Yukawa processes is derived as
\begin{eqnarray}
Y_{\rm DM}^{\nu-\rm Yukawa} &\simeq& \frac{y^2 \left( Y_1^2 + Y_2^2 \right)}{256\pi^5 m_\phi} C \times 5.5\,.
\label{eq:Yieldyuk}
\end{eqnarray}
The yield contributed by the $\nu$-Yukawa processes depends on both the neutrino portal coupling $y$ and the sum of the squared seesaw couplings. If the $\nu$-Yukawa processes are dominating the DM production, the required relation between the seesaw Yukawa couplings and the neutrino portal coupling can be estimated by requiring the yield in Eq.\eqref{eq:Yieldyuk} equals the yield in Eq.\eqref{eq:omegaPREA}. The result is
\begin{eqnarray}
y^2\left( Y_1^2 + Y_2^2 \right) \simeq 2.1 \times 10^{-22} \frac{m_\phi}{m_\chi}\,,
\end{eqnarray}
from which it can be inferred that the relation between the couplings only depends on the ratio of dark particle masses. Although the dark matter production through the $\nu$-Yukawa processes depends on the seesaw couplings, it is not affected by the VEVs of the 2HDM directly as in the cases of muon decay and neutrino mixing. The only influence from the 2HDM parameters is the minimum value of $Y_1^2 + Y_2^2$ in Eq.\eqref{eq:rel3}.

For the dark sector processes, similar treatment can be applied  and the yields follow 
\begin{eqnarray}
Y_{\rm DM}^{\phi\phi} 
\simeq \frac{y^4}{256\pi^5 m_\phi} C \times 19.6  \,
\quad \text{and} \quad 
Y_{\rm DM}^{\chi\chi} \simeq \frac{y^4}{256\pi^5 m_\phi} C \times 9.8\,.
\end{eqnarray}
The ratio of yields from $\phi\phi$ and $\chi\chi$ process is around $2$, which can be read from the spin-averaged amplitudes at high energy limit $\overline{\left|\mathcal{M}\right|^2}_{\phi\phi^*\rightarrow N\overline{N}}\simeq 2\overline{\left|\mathcal{M}\right|^2}_{\chi\overline{\chi}\rightarrow N\overline{N}}\simeq 2y^4$. In total, the contribution from the dark sector processes is
\begin{eqnarray}
Y_{\rm DM}^{\rm DS} \simeq \frac{y^4}{256\pi^5 m_\phi} C \times 29.5\,.
\label{eq:YieldDS}
\end{eqnarray}
When the dark sector coupling is large enough, the dark sector processes can dominate the dark matter production and the coupling is determined by 
\begin{eqnarray}
y^4 \simeq 3.9 \times 10^{-23} \frac{m_\phi}{m_\chi}.
\end{eqnarray}
With Eq.\eqref{eq:Yieldyuk} and Eq.\eqref{eq:YieldDS}, the dominance of dark matter production can be obtained by evaluating the ratio of the yields $r_Y \equiv Y_{\rm DM}^{\rm DS}/Y_{\rm DM}^{\nu-\rm Yukawa}$. If $y/\sqrt{Y_1^2+Y_2^2} > 0.43$, $r_Y > 1$ and the dark sector processes dominate the dark matter production; if $y/\sqrt{Y_1^2+Y_2^2} < 0.43$, $r_Y < 1$ and the $\nu$-Yukawa processes dominate the dark matter production. When $y/\sqrt{Y_1^2+Y_2^2} = 0.43$, $r_Y = 1$ and the contributions from dark sector and $\nu$-Yukawa processes are equal. In the case $r_Y = 1$, the seesaw couplings satisfy 
\begin{eqnarray}
\left( Y_1^2 + Y_2^2 \right)^2 \simeq 5.5 \times 10^{-22} \frac{m_\phi}{m_\chi}\,.
\label{eq:couplingseq}
\end{eqnarray}
As dark scalar decay is required for single dark matter scenario, the ratio of dark particles mass always satisfies $m_\phi/m_\chi>1$ and therefore $Y_1^2 + Y_2^2 \gtrsim 2.3 \times 10^{-11}$. If only one of the seesaw couplings dominates, the minimum value of the dominating coupling is around $4.8 \times 10^{-6}$. Notice that the quantity $Y_1^2 + Y_2^2$ is constrained by Eq.\eqref{eq:rel3} and it can be turned into a constraint on the dark particles mass ratio $m_\phi/m_\chi$ 
\begin{eqnarray}
\frac{m_\phi}{m_\chi}\gtrsim 6.9\times 10^{-9}  \left(\frac{M_N}{1 \text{GeV}}\right)^2 \frac{1}{\sin^2{2\beta}} \,.
\label{eq:minratio}
\end{eqnarray}
When the right side of Eq.\eqref{eq:minratio} is larger than 1, the $\nu$-Yukawa processes dominate the dark matter production definitely when $m_\phi/m_\chi$ is below the threshold value.

\section{Results\label{sec:Results}}

In this section, we show the numerical result for relation between the seesaw couplings and the neutrino portal coupling, and discuss the constraints from existing and future experiments. We use the open code MicrOmegas~\cite{Belanger:2018ccd}, with model generated by LanHEP~\cite{Semenov:2008jy}, to compute the dark matter relic density. As the first step, we start with the relation between the couplings, which can be compared with the analytical calculations. The numerical results show that
\begin{eqnarray}
y^2 \left( Y_1^2 + Y_2^2 \right) \simeq 1.9\times 10^{-22} \frac{m_\phi}{m_\chi},
\end{eqnarray}
when the $\nu$-Yukawa processes dominate the dark matter production and
\begin{eqnarray}
y^4 \simeq 5.1\times 10^{-23} \frac{m_\phi}{m_\chi}.
\end{eqnarray}
when the dark sector processes dominate. These numerical results are consistent with the analytical calculation in Sec.\ref{sec:DMpro}, within the errors from the approximations applied.

\begin{figure}[t!]
\begin{center}
\subfigure[~$M_N=1$ GeV]{\includegraphics[width=0.48\textwidth]{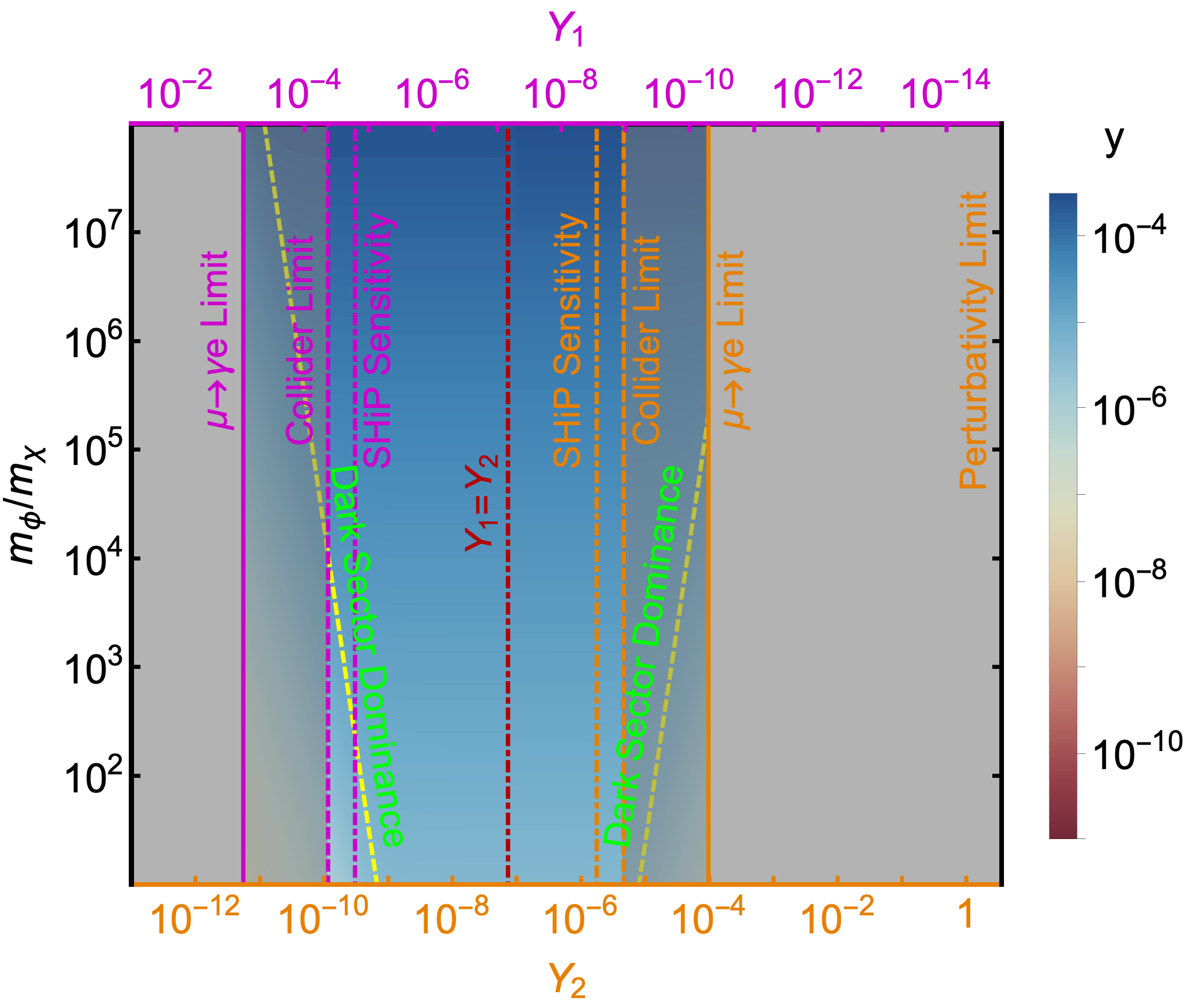}\label{fig:y2rN0}}
\subfigure[~$M_N=10^2$ GeV]{\includegraphics[width=0.48\textwidth]{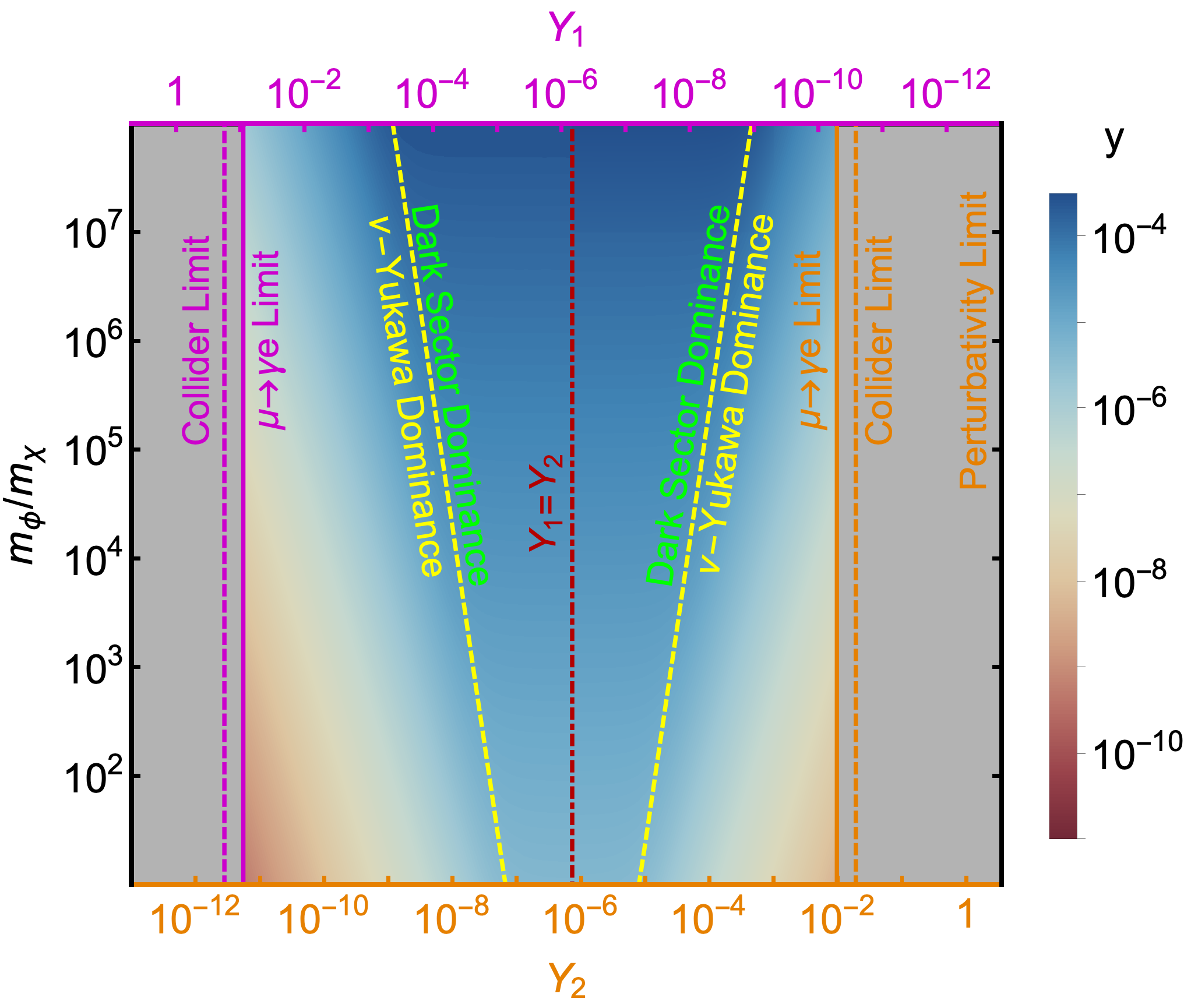}\label{fig:y2rN2}}
\subfigure[~$M_N=10^4$ GeV]{\includegraphics[width=0.48\textwidth]{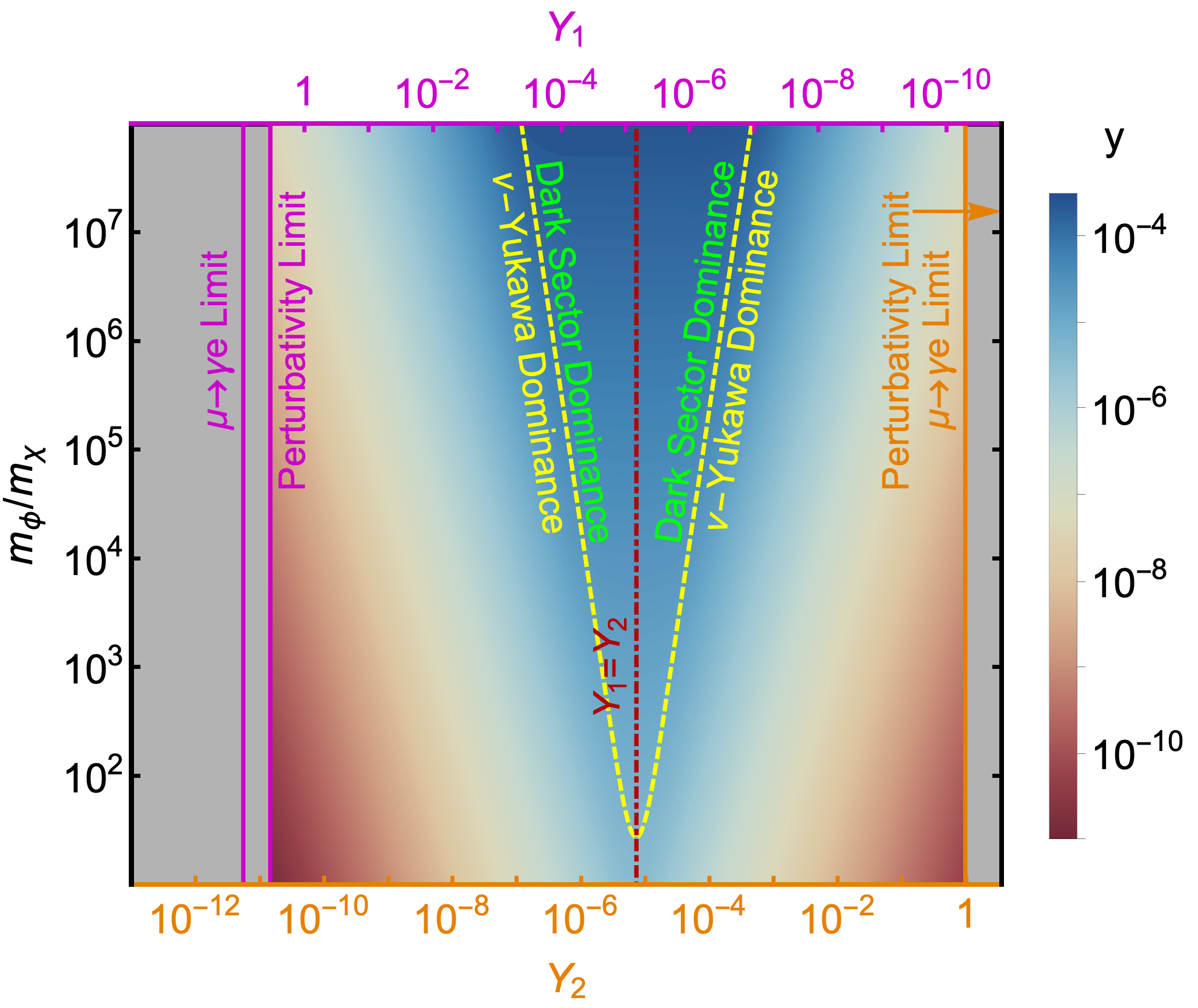}\label{fig:y2rN4}}
\subfigure[~$M_N=10^6$ GeV]{\includegraphics[width=0.48\textwidth]{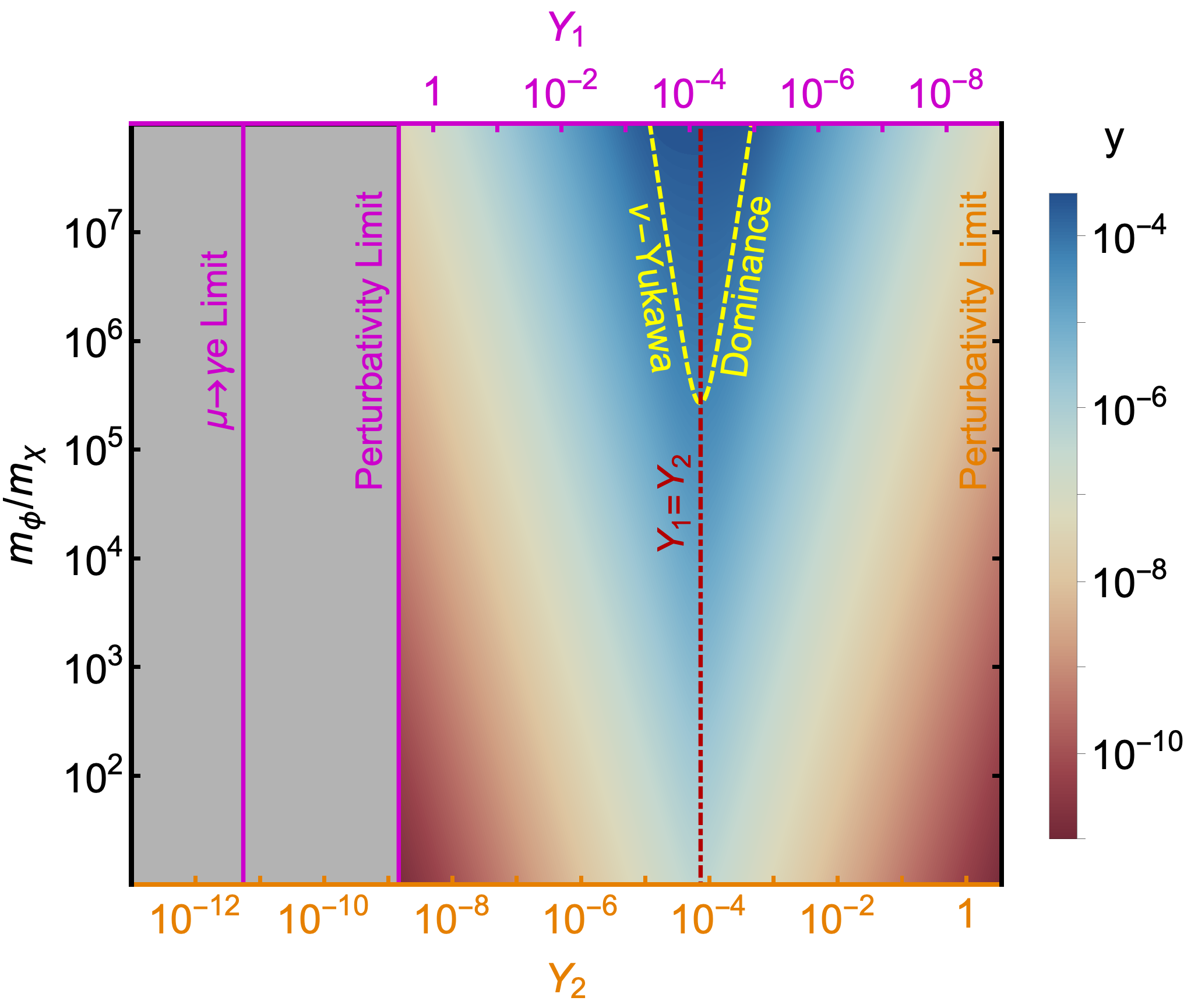}\label{fig:y2rN6}}
\caption{\label{fig:y2r} Required portal coupling $y$ in order to achieve the correct dark matter relic abundance in the $m_\phi/m_\chi - Y_2$ plane with different heavy neutrino mass. The reheating temperature is set to be $T_{\rm RH}=10^{10}$ GeV and $\tan\beta=10$ to avoid gravitational effects~\cite{Chianese:2020khl}. $m_\phi$ is fixed to be $10^8$ GeV and $m_\chi$ changes from $1-10^7$GeV. The relation between $Y_1$ and $Y_2$ is determined by Eq.\eqref{eq:rel2}. The neutrino Yukawa process and dark sector process make equal contribution to the relic abundance along the yellow lines. }
\end{center}
\end{figure}

Following a full numerical calculation, which does not use the above approximations,
the required portal coupling $y$ for the observed relic abundance is shown in Fig.\ref{fig:y2r}. The value of $y$ only depends on the ratio of dark particle masses $m_\phi/m_\chi$ and $Y_1^2+Y_2^2$ if $T_{\rm RH}\gg m_\phi \gg m_\chi,\,M_N$. Therefore the values of $y$ in different panels of Fig.\ref{fig:y2r} are roughly the same for the same values of $m_\phi/m_\chi$ and $Y_1^2+Y_2^2$. The red lines represent $Y_1=Y_2$, which depend on $M_N$ due to Eq.\eqref{eq:rel3}. $Y_1$ dominates the neutrino Yukawa production of DM on the left of the red lines while $Y_2$ dominates on the right. The figures are symmetric relative to the red lines since the dependence is on $Y_1^2+Y_2^2$. The grey areas are excluded either by the muon decay or by the perturbativity limit of the neutrino portal coupling. The light shadowed areas in Fig.\ref{fig:y2rN0} are constrained by the collider data while it is weaker than the constraint from muon decay in Fig.\ref{fig:y2rN2}. The future SHiP sensitivity is also marked out in Fig.\ref{fig:y2rN0} as dot-dashed lines.

In Fig.\ref{fig:y2r}, the yellow lines mark where the $\nu$-Yukawa processes and dark sector processes contribute equally to the relic abundance. The dark sector processes are dominating the dark matter production above the yellow lines while the $\nu$-Yukawa processes are dominating under them. Along the yellow lines, the ratio $y/\sqrt{Y_1^2+Y_2^2}$ has numerical values around $0.52$, agreeing with the analytical result. In Fig.\ref{fig:y2rN0}, where the parameter space is highly constrained by the collider data, the dark sector dominance is favoured: the $\nu$-Yukawa dominance is fully constrained when $Y_2$ dominates the seesaw couplings and only a small parameter space is allowed for $\nu$-Yukawa dominance when $Y_1$ dominates. In Fig.\ref{fig:y2rN4} and Fig.\ref{fig:y2rN6}, there are minimum values of $m_\phi/m_\chi$ for the dark sector process to dominate. The reason is that the relation in Eq.\eqref{eq:couplingseq} cannot be satisfied for small $m_\phi/m_\chi$ since the minimum value of $Y_1^2+Y_2^2$ increases as the heavy neutrino becomes massive in Eq.\eqref{eq:rel3}. As a result, the dark matter production is definitely dominated by $\nu$-Yukawa processes when $m_\phi/m_\chi$ is below the threshold value determined by the heavy neutrino mass. In particular, the dark matter production cannot be dominated by the dark sector process when $m_\phi/m_\chi$ is smaller than 28 and $2.8\times10^5$ in Fig.\ref{fig:y2rN4} and Fig.\ref{fig:y2rN6}, respectively, and the results agree with Eq.\eqref{eq:minratio}.

{In general, the DM detection can also constrain the neutrino portal coupling $y$. It can be derived that the sensitivity of the next generation experiment Hyper-Kamiokande (HK) \cite{Abe:2018uyc} on relic density cross section ($\left<\sigma v\right> =3 \times 10^{-26}\, \text{cm}^3/\text{s}$) roughly corresponds to $y^2 \lesssim (m_\chi^2 + m_\phi^2)/( m_\chi \times 1 \text{GeV}$) \cite{Blennow:2019fhy}. Such constraint is stronger than the perturbativity limit of $y$ only if the dark scalar $\phi$ is not much heavier than the dark fermion $\chi$ as well as $1$ GeV. In our model, as shown at the bottom of Fig.\ref{fig:y2rN0}, the portal coupling $y$ is roughly smaller than $10^{-5}$ when both requirements are best satisfied, while  the sensitivity of the next generation experiment is larger than 1. Therefore the DM detection bound does not constrain the free parameters in the model.}

\begin{figure}[t!]
\begin{center}
\includegraphics[width=0.48\textwidth]{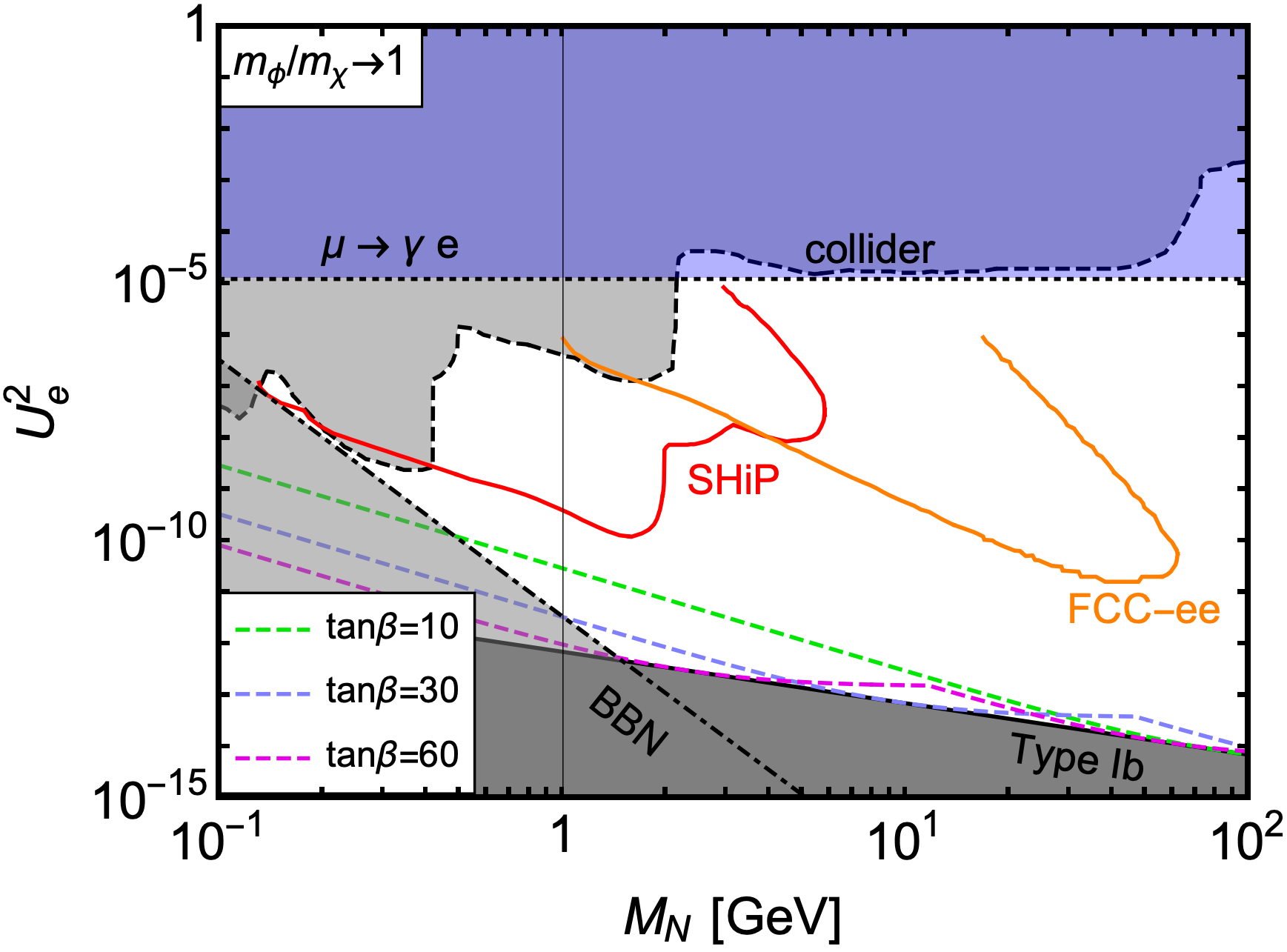}
\includegraphics[width=0.48\textwidth]{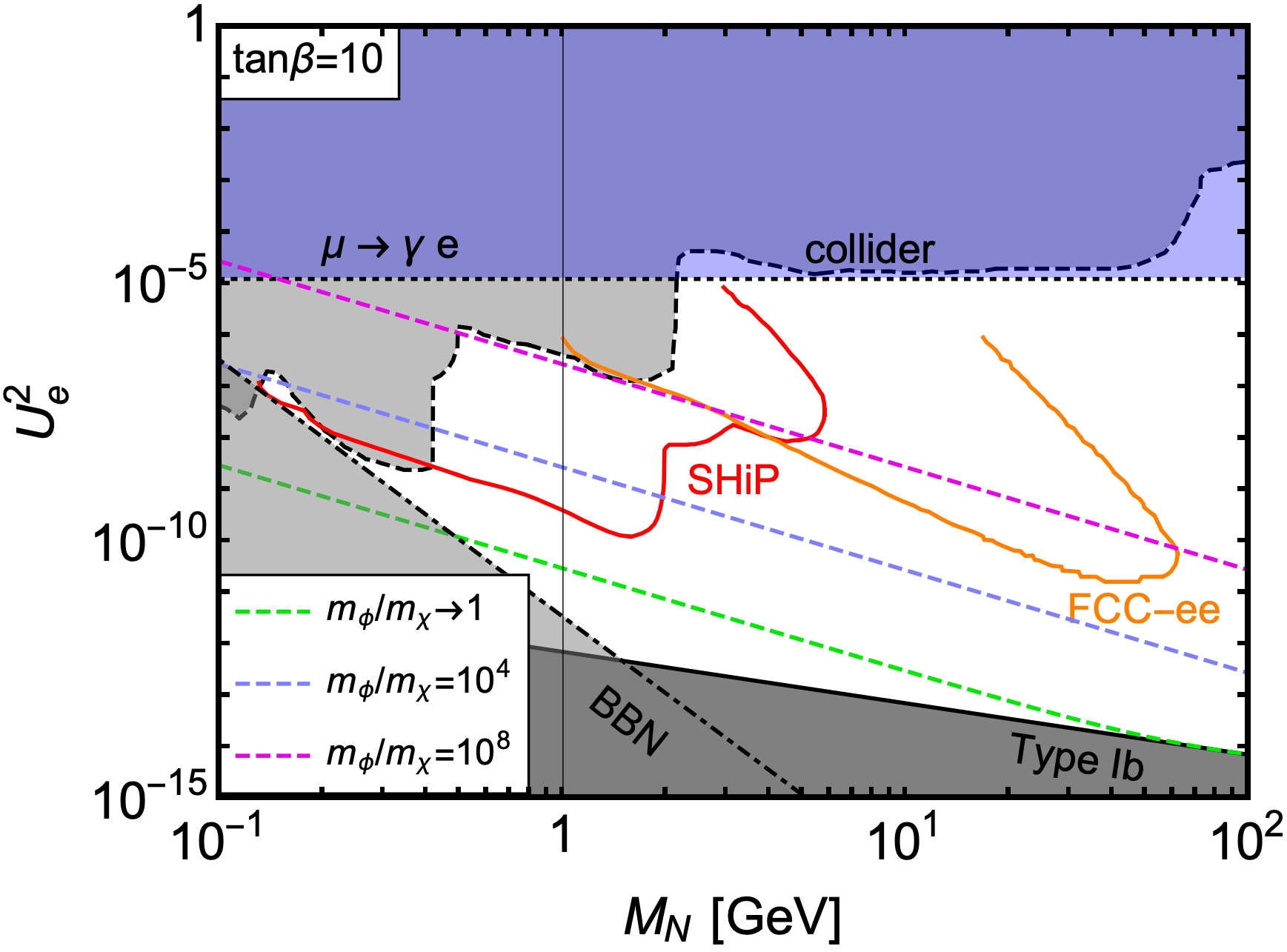}
\includegraphics[width=0.48\textwidth]{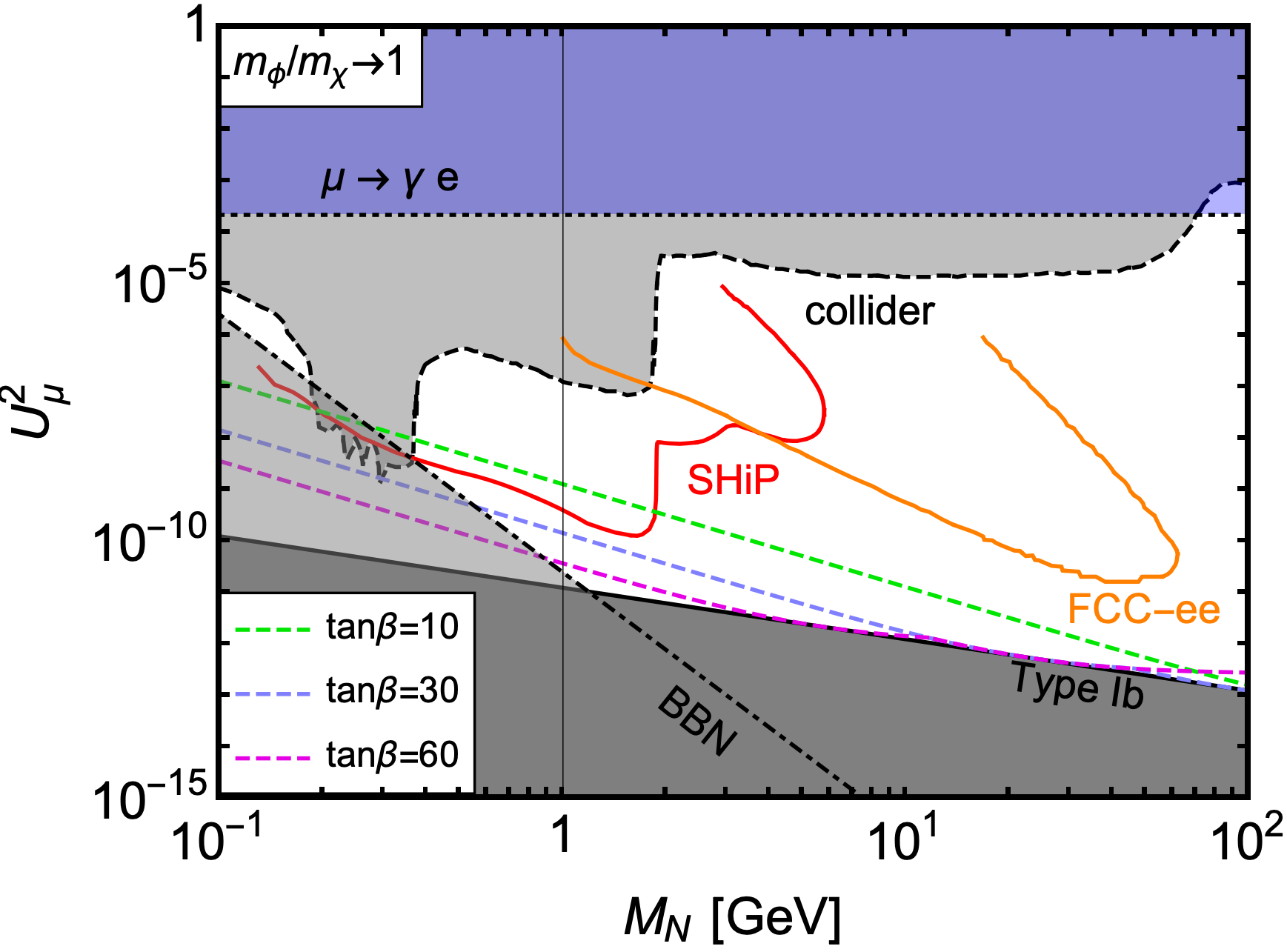}
\includegraphics[width=0.48\textwidth]{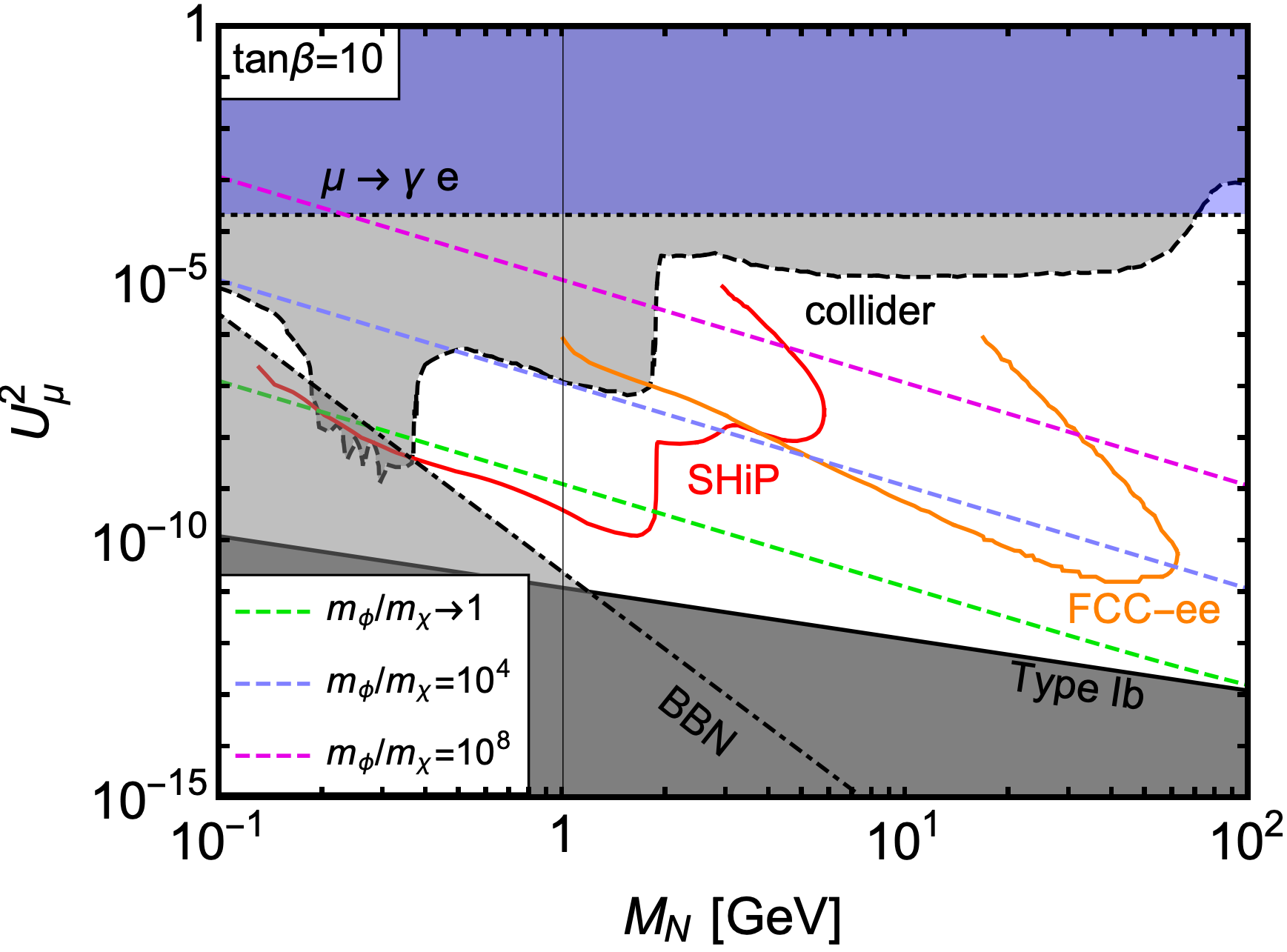}
\includegraphics[width=0.48\textwidth]{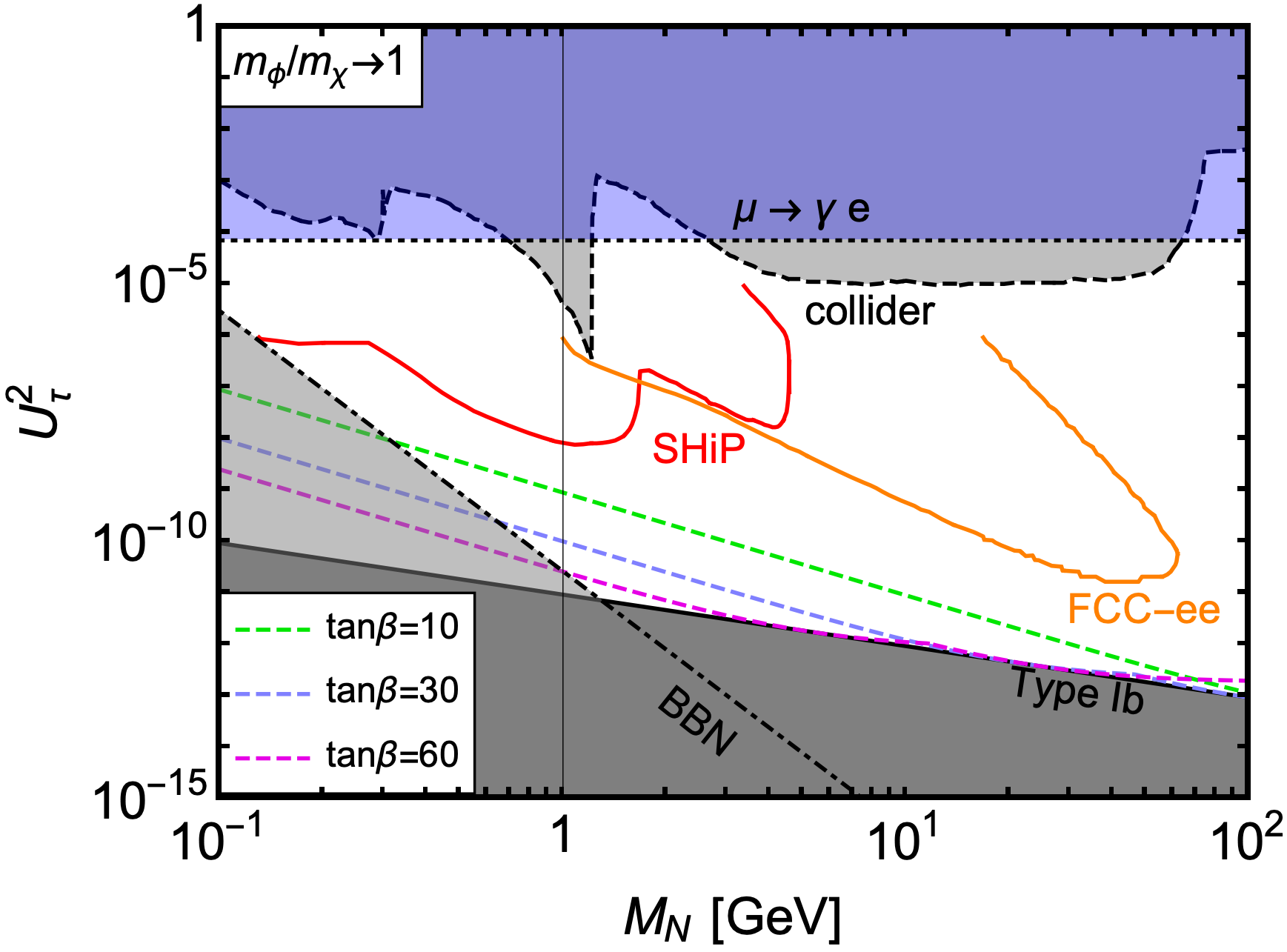}
\includegraphics[width=0.48\textwidth]{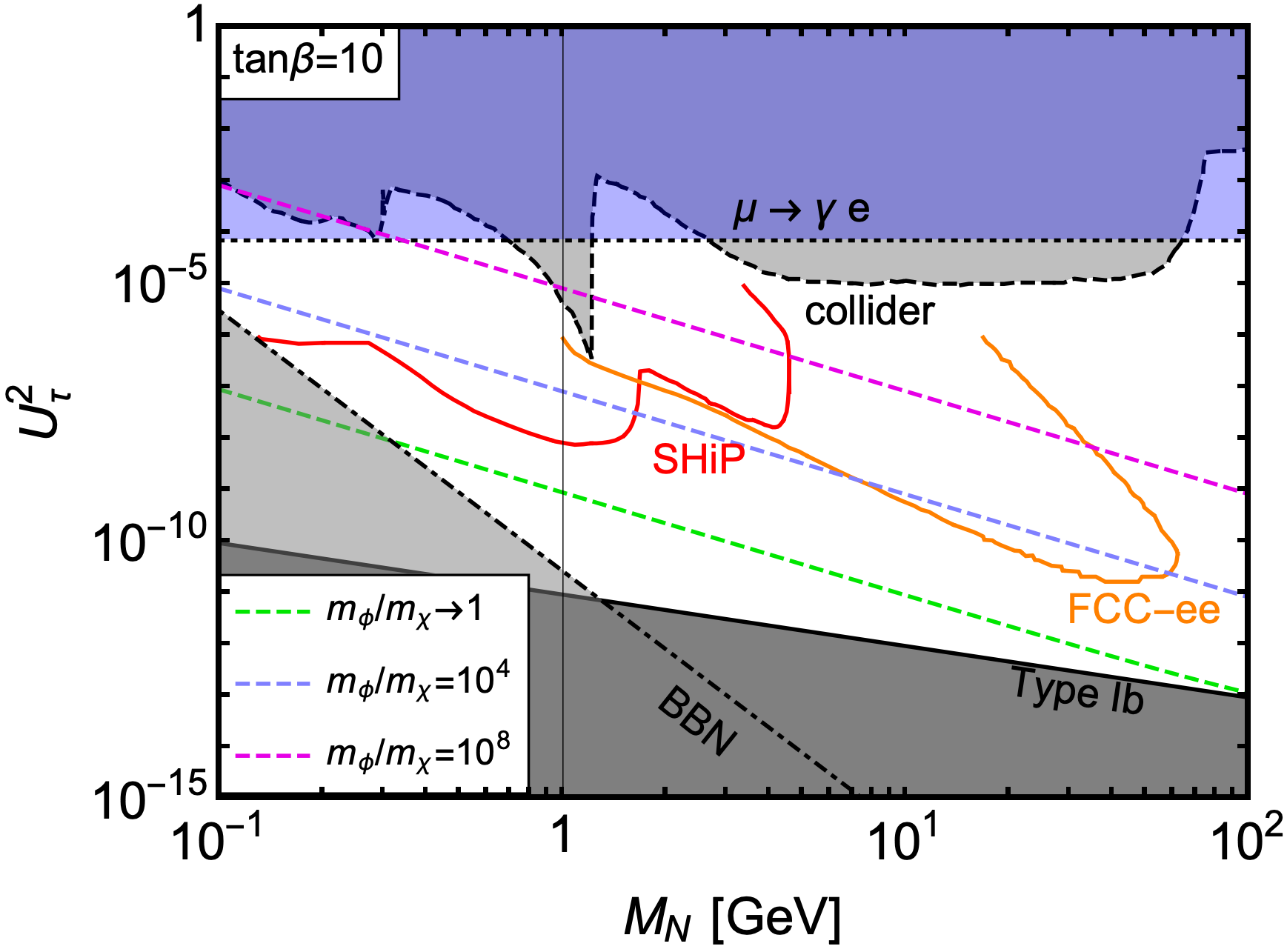}
\caption{\label{fig:U2}  Constraints and predicted dark matter dominance in the $M_N$-$U_\alpha^2$ plane of the minimal type Ib seesaw model in which there is only a single heavy Dirac neutrino of mass $M_N$. The correct relic abundance of dark matter can be produced over the entire white region. The left panels show the regions of definite dark sector dominance (meaning that the neutrino Yukawa couplings definitely do not play a role in dark matter production), which occurs in the white regions below the  coloured dashed lines, for different values of $\tan\beta$. The right panels show analogous regions for different ratios of dark particle masses. The black lines mark the constraints on the quantity $U_\alpha^2$ in the type Ib model. The red and orange lines stand for the future sensitivity of SHiP~\cite{SHiP:2018xqw} and FCC-$ee$~\cite{Blondel:2014bra}. }
\end{center}
\end{figure}

Fig.\ref{fig:U2} shows the constraints and predicted dark matter dominance. As many experimental constraints are directly linked to the neutrino mixing, the result is presented in $M_N$-$U_\alpha^2$ plane, where $U_\alpha^2$ are defined by Eq~\eqref{eq1:U2} in the framework of Type Ib seesaw model. The black lines show the constraints on $U_\alpha^2$. Below the solid black line, the parameter space is excluded by the structure of Yukawa couplings as shown in Eq.\eqref{eq3:U2}, which is determined by the neutrino data, regardless of the Majorana phase $\delta_M$. The region above the dotted line is excluded by the muon decay. Although both the muon decay constraint and the expression of $U^2$ depend on the heavy neutrino mass, the dependence is cancelled since both of them are proportional to $v_i^2 Y_i^2/M_N^2$ when only one of the Yukawa couplings dominates, and therefore muon decay constraint appears to be independent of $M_N$. The shadowed region above the dashed line is excluded by the collider data from multiple experiments~\cite{Deppisch:2015qwa} and the one below the dash-dotted line is excluded by Big Bang Nucleosynthesis (BBN) data~\cite{Canetti:2012kh,Drewes:2019mhg}.\footnote{Within the range of heavy neutrino mass considered in Fig.\ref{fig:U2}, the perturbativity limit is always weaker than the constraint from collider data and the muon decay, as indicated by Fig.\ref{fig:y2rN0} and Fig.\ref{fig:y2rN2}. } Besides the existing constraints, the future experiment sensitivity of SHiP~\cite{SHiP:2018xqw} and FCC-$ee$~\cite{Blondel:2014bra} are also shown in Fig.\ref{fig:U2} as the red lines and the orange lines, respectively. 

The green, blue and purple lines in Fig.\ref{fig:U2} mark the lowest values of $U_\alpha^2$ that the $\nu$-Yukawa process can dominate the dark matter production for different benchmarks, i.e. the dark matter production is definitely dominated by dark sector process below these lines for the corresponding benchmarks. Along those lines, the Yukawa couplings are fixed to the threshold values corresponding to the selected dark particle mass ratios on the yellow lines in Fig.\ref{fig:y2r}.

In the left panels, the dashed coloured lines are obtained in the limit $m_\phi/m_\chi\rightarrow1$ which can never be actually reached due to the required mass ordering $m_\phi > m_\chi + M_N$ in the single dark matter scenario. In such a limit, the dominance of dark matter production switches when the dominant coupling is around $4.5 \times 10^{-6}$. The colours green, blue and purple stand for $\tan\beta$ equals 10, 30 and 60, which cover the maximum value of $\tan\beta$ in most recent literatures~\cite{Degrande:2016hyf,Akeroyd:2016ymd,Misiak:2017bgg,Ivanov:2017dad,Arbey:2017gmh,Aaboud:2017gsl,Aaboud:2018gjj,Sirunyan:2018koj,Aad:2020fpj,Chen:2020aht,Su:2020pjw,Bertrand:2020lyb}. These lines show interesting behaviours as $\tan\beta$ and $M_N$ change. To understand their behaviours, the first step is to notice that $U_\alpha^2$ are always determined by their values when the dark matter production is driven by $Y_1$. Suppose the value of $Y_1$ ($Y_2$) when the dominance of dark matter production switches in the $Y_1$ ($Y_2$) dominating region is $Y_1^0$ ($Y_2^0$). Then $Y_1^0=Y_2^0$ since the figures in Fig.\ref{fig:y2r} are symmetric relative to the red lines $Y_1=Y_2$. And as shown in Eq.~\eqref{eq2:U2}, $U_\alpha^2$ are proportional to either $v_1Y_1$ or $v_2Y_2$ when their ratio is far from one, with the same coefficients. However, in general, $Y_1$ dominance in dark matter production does not mean $v_1Y_1$ dominance in $U_\alpha^2$. Indeed, according to Eq.~\eqref{eq:rel1}, $v_1Y_1$ dominance requires $Y_1^2>3.0\times 10^{-11} \,\text{GeV} M_N/v_1^2$ while $Y_1$ dominance requires $Y_1^2>3.0\times 10^{-11} \,\text{GeV} M_N/(v_1v_2)$. When $M_N$ is larger than $(Y_1^0)^2v_1^2/(3.0\times 10^{-11} \,\text{GeV})$, $v_2Y_2$ dominates $U_\alpha^2$ while $Y_1$ dominates the dark matter production. This situation does not appear when $Y_2$ dominates the dark matter production as $\tan\beta > 1$. Therefore the dark sector dominance and $U_\alpha^2$ dominance regarding to the Yukawa couplings have three scenarios: (1) $Y_1$ and $v_1Y_1$ (2) $Y_1$ and $v_2Y_2$ (3) $Y_2$ and $v_2Y_2$. $Y_1=Y_1^0$ in scenario (1) and (2) while $Y_2=Y_2^0$ in scenario (3). From the discussion before, it is easy to conclude that the $U_\alpha^2$ in scenario (1) is smaller than the $U_\alpha^2$ in scenario (3). In scenario (2), $U_\alpha^2$ is proportional to $v_2Y_2$ which is smaller than $v_2Y_1^0$, and therefore smaller than the value of $U_\alpha^2$ in scenario (3). In summary, $U_\alpha^2$ are always determined by the scenario when $Y_1$ dominates the dark matter production.

As a result, the dashed coloured lines move downwards as $\tan\beta$ increases when the heavy neutrino is light, because $U_\alpha^2$ are proportional to $v_1^2Y_1^2$ and thus $\cos^2\beta$ in that region. It can be observed that those lines tend to touch the type Ib limit as the heavy neutrino mass increases. The reason for such tendency is because the values of the Yukawa couplings when $U_\alpha^2$ is minimised is proportional to $\sqrt{M_N}$ as shown in Eq.~\eqref{eq:U2minY}. After $Y_1$ become smaller than the value for minimum $U_\alpha^2$, the lines leave the type Ib limit. One may observe, especially in the case of $\nu_e$, that some of the lines approach the type Ib limit again after leaving it. This is because $U_\alpha^2$ become proportional to $v_2^2Y_2^2$ rather than $v_1^2Y_1^2$ as the mass of the heavy neutrino grows. In the case of $m_\phi/m_\chi\rightarrow1$, this change happens when $M_N$ is larger than $4.2\times10^4\,\text{GeV} \cos^2\beta$ (around 47 and 12 GeV for $\tan=$  30 and 60). As $\tan\beta$ grows, the region for definite dark sector dominance becomes small, which means the $\nu$-Yukawa dominance is less constrained. 

In the right panels, the green, blue and purple lines stand for the mass ratio of dark particles 1, $10^4$ and $10^8$, with $\tan\beta=10$. As $U_\alpha^2$ are proportional to $v_1^2Y_1^2$ and thus $\sqrt{m_\phi/m_\chi}$ when $Y_1$ dominates the dark matter production according to Eq.~\eqref{eq:couplingseq}, the lower limit of $U_\alpha^2$ for $\nu$-Yukawa dominance move up as the mass ratio of dark particles increases and the intervals between two adjacent lines are roughly two orders of magnitude. In the case $M_N=1$ GeV, the $\nu$-Yukawa dominance is almost forbidden due to collider constraint in $\nu_\mu$ mixing for $m_\phi/m_\chi=10^4$ and totally excluded for $m_\phi/m_\chi=10^8$, which is consistent with the result in Fig.\ref{fig:y2rN0}. The larger mass ratio $m_\phi/m_\chi$ is, the easier the $\nu$-Yukawa dominance is to be tested by the upcoming SHiP and FCC-$ee$ results.

\section{Conclusion \label{sec:Concl}}
In this paper, we have proposed a minimal type Ib seesaw model, based on type II 2HDM,
where a $Z_3$ symmetry ensures that the effective neutrino mass operator involves 
two different Higgs doublets, and the two right-handed neutrinos form a single heavy Dirac mass,
providing an extremely simple seesaw model with low scale testability. For example the whole of neutrino mass and mixing may be accounted for by a single heavy Dirac neutrino of mass around the GeV scale with large couplings to SM fermions, making it eminently discoverable at colliders and SHiP. We emphasise that there are no other heavy neutrinos required, since only one Dirac neutrino is needed in the minimal model. To explain dark matter,  
we have discussed a minimal extension of this model, in which the single heavy Dirac neutrino of mass $M_N$
is coupled to a dark Dirac fermion and a dark complex scalar field, both charged under a discrete $Z_2$ symmetry, where the lighter of the two, assumed to be the fermion, is a dark matter candidate. It is remarkable that such a single heavy Dirac neutrino of mass around the GeV scale can not only account for neutrino mass and mixing but can also act as a portal for dark matter. 

We have studied analytically and numerically the dark matter production in the simple dark matter extension of the minimal type Ib seesaw model, with the heavy Dirac neutrino portal to a dark scalar and a dark fermion. Due to the special structure of the type Ib seesaw model, the parameters in the model are highly constrained by the oscillation data and the dark matter production has an interesting dependence on the seesaw Yukawa couplings. It has been proved analytically and confirmed numerically that the dark matter production only depends on the ratio of dark particle masses in the case of non-degenerate masses. We have shown the required neutrino portal coupling for different values of seesaw couplings and dark particle mass ratio and highlighted the regions for different production mechanism. Since the dark matter production has a symmetric dependence on the seesaw couplings, the required value of portal coupling is symmetric with respect to the two seesaw couplings. Although dark matter production involving the type Ib seesaw Yukawa interaction is favoured when the mass of the heavy neutrino is large, it is still constrained for the GeV mass heavy neutrino accessible to low energy experiments. 

We have presented the regions of parameter space where the dark matter can be produced through the type Ib seesaw Yukawa interaction with the neutrino mixing characterised by the quantities $U_\alpha^2$ which are relevant for experiment. The regions where the type Ib seesaw Yukawa interaction can affect dark matter production are shown for different benchmark values of $\tan\beta$ and dark particle mass ratios. In all allowed regions dark matter may be produced through the Dirac neutrino portal.
The neutrino mixing is seen to be constrained by the current experimental results and is testable at future experiments such as SHiP and FCC-$ee$, especially when $\tan\beta$ is small and the dark particle mass ratio is large. The discovery of the single heavy Dirac neutrino of the minimal type Ib seesaw model would not only unlock the secret of the origin of neutrino mass but could also provide important information on the mechanism of dark matter production.

\acknowledgments

MC acknowledges partial support from the research grant number 2017W4HA7S “NAT-NET: Neutrino and Astroparticle Theory Network” under the program PRIN 2017 funded by the Italian Ministero dell’Università e della Ricerca (MUR) and from the research project TAsP (Theoretical Astroparticle Physics) funded by the Istituto Nazionale di Fisica Nucleare (INFN). BF acknowledges the Chinese Scholarship Council (CSC) Grant No.\ 201809210011 under agreements [2018]3101 and [2019]536. SFK acknowledges the STFC Consolidated Grant ST/L000296/1 and the European Union’s Horizon 2020 Research and Innovation programme under Marie Sklodowska-Curie grant agreement HIDDeN European ITN project (H2020-MSCA-ITN-2019//860881-HIDDeN).

\appendix

\section{Constraints on the Higgs portal couplings}

In principle, the model also allows the direct coupling between the Higgs doublets and the dark scalar. The corresponding Lagrangian is 
\begin{eqnarray}
\mathcal{L}_{\rm Higgs Portal} =\lambda_{\Phi\phi} |\phi|^2|\Phi_1|^2 + \lambda_{\Phi\phi}' |\phi|^2|\Phi_2|^2\,.
\label{eq:Higgs} 
\end{eqnarray}
As analysed in \cite{Chianese:2020yjo} and mentioned in \cite{Chianese:2020khl}, the Higgs portal production is proportional to the squared couplings $\lambda_{\Phi\phi}^2\,,\lambda_{\Phi\phi}'^2$ and the ratio of dark particle masses $m_\phi/m_\chi$. In this model, numerical results show that 
\begin{eqnarray}
{\Omega_{\rm DM} h^2}_{\rm Higgs Portal} \simeq 7.2 \times 10^{20} \left(\lambda_{\Phi\phi}^2+\lambda_{\Phi\phi}'^2\right)\frac{m_\chi}{m_\phi}
\label{eq:HiggsRA} 
\end{eqnarray}
In order to make the dark matter produced by Higgs portal effects contribute less than 10\% of the observed relic abundance, the couplings have to satisfy 
\begin{eqnarray}
\sqrt{\lambda_{\Phi\phi}^2+\lambda_{\Phi\phi}'^2} \lesssim 4.1 \times 10^{-12} \sqrt{\frac{m_\phi}{m_\chi}}\,.
\label{eq:HiggsCon} 
\end{eqnarray}
As we intend to focus on the relation between the neutrino physics and dark matter, we assume the Higgs portal coupling is always small enough to make a neglectable contribution to dark matter production.

\bibliographystyle{JHEP}
\bibliography{Type1b_bib}

\end{document}